\documentclass[3p]{elsarticle}
\usepackage[ruled, linesnumbered]{algorithm2e}
\usepackage{caption}
\usepackage{subcaption}
\usepackage{hyperref}
\usepackage{xcolor}
\usepackage{float}
\usepackage{multirow}
\usepackage{amsmath}
\usepackage{booktabs}
\usepackage{dsfont}
\usepackage[toc,page]{appendix}

\title{Explainable Malware Detection through Integrated Graph Reduction and Learning Techniques}

\begin{document}

\begin{frontmatter}

\author[1]{Hesamodin Mohammadian \corref{cor1}}
\ead{h.mohammadian@unb.com}

\author[1]{Griffin Higgins}
\ead{griffin.higgins@unb.ca}

\author[1]{Samuel Ansong}
\ead{samuel.ansong@unb.ca}

\author[1]{Roozbeh Razavi-Far}
\ead{roozbeh.razavi-far@unb.ca}

\author[1]{Ali A. Ghorbani}
\ead{ghorbani@unb.ca}

\affiliation[1]{organization={Canadian Institute for Cybersecurity, University of New Brunswick},
city={Fredericton, New Brunswick},
country={Canada}}

\cortext[cor1]{Corresponding author}

\begin{abstract}
Control Flow Graphs and Function Call Graphs have become pivotal in providing a detailed understanding of program execution and effectively characterizing the behavior of malware. These graph-based representations, when combined with Graph Neural Networks (GNN), have shown promise in developing high-performance malware detectors. However, challenges remain due to the large size of these graphs and the inherent opacity in the decision-making process of GNNs. This paper addresses these issues by developing several graph reduction techniques to reduce graph size and applying the state-of-the-art GNNExplainer to enhance the interpretability of GNN outputs. The analysis demonstrates that integrating our proposed graph reduction technique along with GNNExplainer in the malware detection framework significantly reduces graph size while preserving high performance, providing an effective balance between efficiency and transparency in malware detection.
\end{abstract}


\end{frontmatter}

\section{Introduction}
In recent years, the threat of malware attacks has increased significantly up to more than six billion in 2023~\cite{CyberArk}. With the requirement to analyze and detect this huge amount of attacks, the traditional malware detection techniques including signature-based methods seemed insufficient. Despite their fast detection and widespread usage, they are unable to detect zero-day attacks and advanced malware~\cite{li2019machine}. Due to the ineffectiveness of the conventional signature-based approaches, such as pattern matching, incorporating machine learning (ML) and deep learning (DL) has gotten more attention recently. These methods show superior detection rates, while adapting to new threats and reducing false positives. Several works have focused on ML-based~\cite{pham2018static, khammas2020ransomware, zhang2016malware} or DL-based~\cite{manavi2020new, frederick2022corpus, li2022novel, bensaoud2024cnn} approaches for detecting and classifying malware.

With the increased complexity of new malware samples and the usage of state-of-the-art techniques to bypass detection mechanisms, the need for more effective representation techniques of malware samples has arisen.  More recent studies have proven the superiority of the graph-based structures to characterize and present the malware behavior and help in achieving higher performance for malware detection. Their value lies in offering a comprehensive view of program execution, enabling analysts to grasp program logic, detect vulnerabilities, and uncover malicious activities, such as concealed or obfuscated codes. Due to this, several graph structures such as Control Flow Graph (CFG)~\cite{dovom2019fuzzy, liu2020multifamily, sun2021effective, abusnaina2021dl}, Function Call Graph (FCG)~\cite{zhang2020exploring, cai2021learning, wu2023iot}, and API Call Graph (ACG)~\cite{amer2020dynamic, amer2021multi, li2022dmalnet} are used to present the inputs for DL models, tasked with malware detection. Although several works used graph-based structures with CNN or RNN models, the emergence of graph neural networks with different architectures such as Graph Convolutional Networks (GCN), Graph Isomorphism Networks (GIN), and GraphSage provided better results in combination with CFGs and FCGs.

Despite their potential, several challenges persist in using graph-based structures and learning methods for malware detection. One major challenge is the sheer size and complexity of the information within these graphs, which can hinder the performance and efficiency of graph-learning models. For example, some statically generated graphs, like CFGs and FCGs, may contain millions of nodes, significantly slowing down the learning process. Since these graphs especially CFGs represent every step of an application, the number of lines of codes, multiple branching and nested loops, and library dependencies can be the reason behind the size and the complexity of these graphs. Additionally, the presence of noisy or irrelevant information in these graphs can further degrade the model performance.

To address this, incorporating graph reduction or pruning techniques is essential to the efficiency and performance of the malware detection models. To the best of our knowledge, this is the first work to include graph pruning as a pre-processing step for the malware detection task. We compared several graph reduction techniques from both graph size and detection performance point of view and proposed a novel technique, called Leaf Prune, which considerably improves the performance and efficiency of the malware detection task.

Another challenge lies in the complexity of graph-based models, which often lacks transparency in their decision-making processes—an issue that is particularly critical for malware detection. To mitigate this, various explainability techniques have been proposed for Graph Neural Network (GNN)-based models, aiming to identify the most influential sub-graphs affecting the model's output. To provide more insights on the decisions made, a post-processing step has been designed along with the GNN-based malware detection system. In this respect, we used GNNExplainer, a state-of-the-art explainability technique to find the most important subgraphs and compare its accuracy with those obtained through pruning.

In this paper, we propose a framework that aims to enhance the performance of GNN-based models for malware detection by integrating a graph reduction module into the learning process. Additionally, to provide experts with deeper insights into the model's decisions, we include an explainability module to identify the most significant sub-graphs for each input sample. To ensure a comprehensive evaluation, we employed both CFGs and FCGs with two different node embeddings as inputs to the proposed framework. In summary, the contributions of this article are as follows:

\begin{itemize}
    \item Proposing a comprehensive malware detection framework that includes graph reduction and explainability. The graph reduction improves the efficiency and accuracy of the graph learning process by reducing extremely large and complex graphs, while maintaining a superior performance.
    \item Extract FCGs and CFGs of the benign and malware samples using static analysis.
    \item Use two novel node feature embedding techniques with the goal of embedding function names from FCGs and assembly instruction from CFGs.
    \item Incorporate a graph reduction module in the malware detection framework for the first time and propose a novel graph pruning technique, called Leaf Prune.
    \item Using GNNExplainer as a post-processing step to provide interpretation for the detection system's decisions.
\end{itemize}

In the rest of the paper, first, the background is reviewed in Section~\ref{sec:background}. In Section~\ref{sec:framework}, the proposed framework is explained in detail and the results and analysis are presented in Section~\ref{sec:result}. Finally, in Section~\ref{sec:conclusion}, we have our conclusion and future works.
\section{Background}\label{sec:background}
The increasing complexity and sophistication of malware have necessitated the development of advanced detection methods that can effectively analyze and identify malicious software. Traditional signature-based approaches, while effective against known threats, often fail to detect novel or obfuscated malware. This limitation has led to a shift towards more dynamic and robust methods, such as those based on CFGs and FCGs, in combination with GNNs.

CFGs and FCGs have become central to modern malware detection due to their ability to represent the structural and functional behaviors of software. A CFG is a representation of all paths that might be traversed through a program during its execution. CFGs are particularly useful for understanding the execution flow of a program and identifying abnormal behaviors that may indicate malicious intent. Similarly, Function Call Graphs (FCGs) depict the relationships between different functions within a program, providing insights into the program's modularity and inter-function communication patterns.

The adoption of CFGs and FCGs in malware detection is driven by their ability to capture the intricate behaviors of malware, which often involve complex control flows and function interactions. By leveraging these graph representations, malware analysts can better understand how malware operates, enabling more accurate detection. Recent advancements have integrated these graph-based representations with machine learning techniques, particularly GNNs, to further enhance the detection capabilities~\cite{gao2022malware,nguyen2018auto,herath2022cfgexplainer}.

Graph Neural Networks (GNNs) have emerged as a powerful tool for analyzing graph-structured data, making them well-suited for malware detection using CFGs and FCGs. GNNs are capable of learning complex patterns in graph data by iteratively aggregating and transforming information from a node's neighbors, enabling them to capture the dependencies and interactions within the graph. This ability makes GNNs particularly effective for understanding the relationships between different parts of a program and identifying patterns indicative of malware.

Node embedding, a key aspect of GNNs, plays a crucial role in transforming the graph's nodes into a continuous vector space, where nodes with similar roles or functions are placed closer together. This transformation allows GNNs to leverage these embeddings to make predictions based on the nature of the graph or specific nodes of it. In the context of malware detection, node embeddings can represent different parts of a program (e.g., basic blocks in CFGs or functions in FCGs), allowing GNN to distinguish between benign and malicious behaviors based on the learned embeddings~\cite{gao2021gdroid,bruna2013spectral,kipf2016semi}.

Graph reduction techniques, such as sparsification, coarsening, and condensation, have become increasingly important in the preprocessing of graph data for GNNs~\cite{hashemi2024comprehensive}. These techniques aim to simplify the graph by reducing its size, while preserving its essential structural properties, which is critical for improving the efficiency and scalability of GNN-based models. In malware detection, graph reduction helps in handling large and complex CFGs and FCGs, making it feasible to apply GNNs to real-world datasets.

For example, sparsification reduces the number of edges in the graph, retaining only the most critical connections, which is essential for focusing the GNN's attention on the most relevant parts of the graph~\cite{hu2013survey}. Coarsening involves merging nodes to create a smaller graph, which can significantly reduce computational complexity~\cite{hashemi2024comprehensive}, on the other side, graph condensation methods  compresses the large, complex graph into a concise, synthetic representation that preserves the most essential and discriminative information of structure and features~\cite{li2023attend}. Pruning, such as leaf or component pruning, removes less significant nodes or components, helping to eliminate noise and improving the clarity of the graph's core structure.

Recent studies have highlighted the effectiveness of graph reduction techniques in various graph-oriented tasks. These techniques, which include sparsification, coarsening, and condensation, have been shown to maintain or even improve performance, while significantly reducing the size and complexity of the graph. By focusing on preserving essential structural properties, graph reduction enables the efficient processing of large and complex graphs without sacrificing accuracy, making it a valuable approach in fields such as malware detection, social network analysis, and biological network modeling. These methods have consistently demonstrated their ability to retain the critical information necessary for the task-specific performance, while optimizing computational resources~\cite{razin2023abilitygraphneuralnetworks,hashemi2024comprehensive}. 

Graph-based representations and GNN-based detection models have shown significant promise in the field of malware detection. However, merely providing a binary classification of ``benign" or ``malicious" is insufficient for security experts and malware analysts, who require detailed insights into the decision-making process~\cite{warmsley2022survey}. To address this need, researchers have integrated explainability modules into malware detection techniques. One of the pioneering works in this area provided textual explanations for malware detection outcomes~\cite{arp2014drebin}. Over the past few years, explainability in malware detection has garnered increasing attention, leading to the development of several methods~\cite{kinkead2021towards, ullah2022explainable, alani2023xmal}. As the focus has shifted toward leveraging GNNs for malware detection, researchers have explored the use of GNN-specific explainability techniques such as GNNExplainer~\cite{ying2019gnnexplainer}, PGExplainer~\cite{luo2020parameterized}, and SubgraphX~\cite{yuan2021explainability}, as well as developing new methods specifically tailored to the malware detection domain~\cite{herath2022cfgexplainer}.
\section{Proposed Framework}\label{sec:framework}
\begin{figure*}
    \centering
    \scalebox{0.55}{%
    \includegraphics{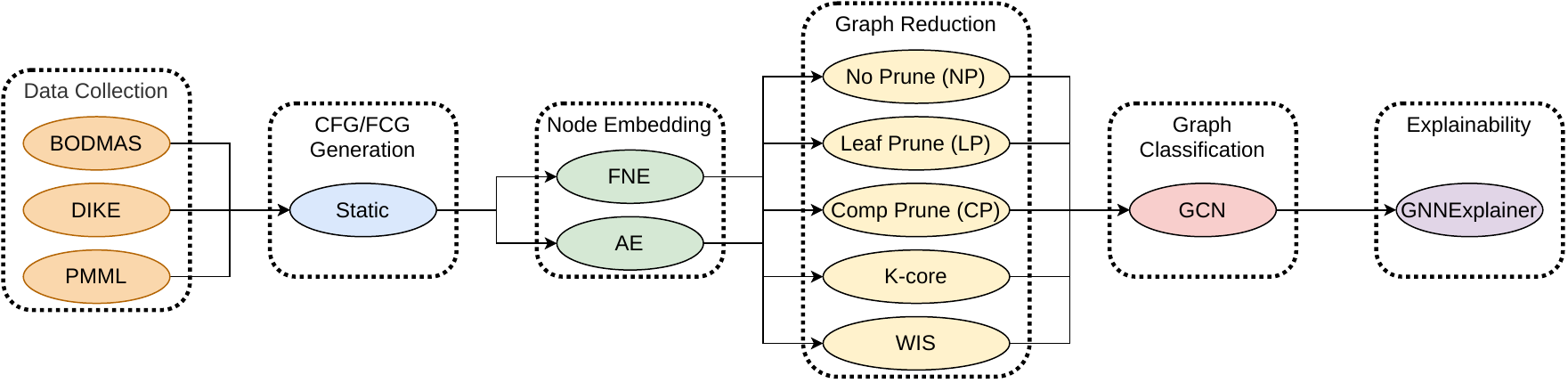}%
    }
    \caption{Proposed framework for malware detection including two modules for data collection and graph generation, two modules for pre-processing, i.e., node embedding and graph reduction, a decision-making module, and a post-analysis module named explainability.}
    \label{fig:pipline}
\end{figure*}

Our proposed malware detection framework extracts graph-based features from the sample through static analysis. After generating CFGs and FCGs of the input samples, two techniques based on sentence transformers and assembly instruction embeddings are used to embed the features of each node of the graphs. In the next step, several graph reduction techniques are used to prune the less important graph's nodes and edges to reduce the graph complexity and increase the detection performance and efficiency. GCN structures are used for the decision-making module. Finally, we have an explainability module using the state-of-the-art GNNExplainer~\cite{ying2019gnnexplainer} technique to find the important sub-graphs of the input sample. Figure~\ref{fig:pipline}, shows the proposed integrated framework, which includes modules and several novel and state-of-the-art techniques.

\subsection{Data Collection}
An essential first stage in the development of machine learning models involves the collection of data that accurately reflects the distribution of binaries observed in real-world circumstances. The improvement of machine learning systems' performance is commonly achieved by raising the quality and amount of labeled data, as supported by known methods such as~\cite{domingos2012few} and~\cite{halevy2009unreasonable}. Nevertheless, the scope of potential binary behaviors is boundless, rendering it impractical to randomly select from the vast array of preexisting binaries. Therefore, it presents a difficulty to determine the complete extent of coverage that our datasets possess throughout the entire range of binary possibilities. The presence of malware in the domain presents further challenges for gathering data, making it impossible to use conventional methods such as using numerous annotators for each file and evaluating their consensus~\cite{geiger2020garbage}.

Surprisingly, malware data is more easily accessible compared to innocuous data. The accessibility of malware samples can be attributed to the presence of platforms that aggregate submissions from volunteers~\cite{Roberts2011, Quist2009} and the implementation of honeypots by researchers~\cite{baecher2006nepenthes}. Conversely, the acquisition of benign data presents much more formidable obstacles. In contrast to malware, benign applications do not aim to spread throughout the Internet, which complicates their collecting process. Thus far, there has been a dearth of scholarly investigations pertaining to the diversity and procurement of benign samples.

Among the plethora of options, BODMAS, Dike, and PMML datasets stand out due to their comprehensive nature and relevance to current cybersecurity challenges. We provide more details about these datasets in Section~\ref{sec:result}.

\begin{figure*}
    \centering
    \scalebox{0.45}{%
    \includegraphics{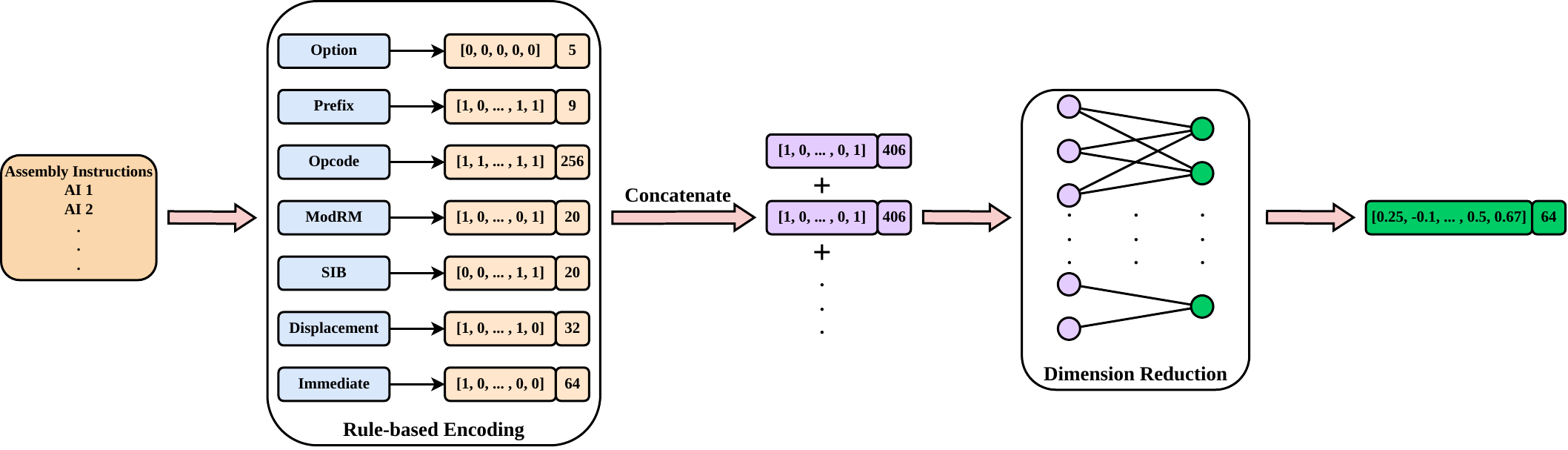}%
    }
    \caption{Schematic diagram of assembly instruction embedding.}
    \label{fig:ae}
\end{figure*}

\subsection{CFG/FCG Generation}
In our work, we utilize the angr~\cite{shoshitaishvili2016state, stephens2016driller, shoshitaishvili2015firmalice} binary analysis Python library for the static recovery of both CFGs and FCGs. Given a binary file as input to angr the static output contains both CFG, FCG, and additional objects that nodes in the respective graphs reference. Importantly, angr uses a combination of both symbolic execution and constraint solving to build graphs. Nodes in CFGs are represented as basic blocks or non-branching sequences of instructions that contain only one entry and exit point. Nodes in FCGs represent functions that contain information such as function names, and other data. Typically, we observe that the number of nodes, edges, and components increases in static CFGs as opposed to FCGs, as observed in Figure~\ref{fig:graph_prune}. This is largely because functions may be composed of many basic blocks, and hence are more condensed by nature. This can be observed in Figure~\ref{fig:vis} where the CFG and FCG of the same binary are compared in the typical malicious and benign case.

\subsection{Node Embedding}
The statically generated CFG/FCG includes a wide range of features that might not be useful and redundant for the GNN-based decision-making process. In this step, we need to decide what features will be used and how they will be represented. We chose function names from the FCGs and assembly instructions from CFGs as the node features that need to be embedded. In the following subsection, we will explain the process of embedding the selected features using state-of-the-art techniques.

\subsubsection{Function Name Embedding (FNE)}
To embed the function names the best choice is to use a pre-trained language model that can map strings to a dense vector space. MiniLM~\cite{wang2020minilm} model is a large language model released by Microsoft that maps paragraphs and sentences to a 348-dimensional space. It has 33 million parameters and has been trained using 1 billion input samples. We fed the function name of each node to this model to get the 384-dimensional output vector and store it as the node feature from now on.

\subsubsection{Assembly Embedding (AE)}
To encode the assembly instruction of the CFG nodes we use the technique presented in~\cite{peng2024malgne} with minor modifications. This technique includes two main steps: Rule-based instruction encoding and machine learning-based dimensionality reduction. Figure~\ref{fig:ae} presents the overall overview of the assembly instruction embedding pipeline.

Each assembly instruction in X86-64 can have up to six components including prefix, opcode, ModRM, SIB, displacement, and immediate. Prefix, opcode, ModRM, and SIB are discrete values and we use one-hot encoding to encode them.

Prefix itself comprises the ES segment register, operand-size override, address-size override, and lock prefix. ES has seven possible values and the other three have two possible values. Therefore, we can one-hot encode the prefix using a nine-dimensional vector.

The opcode is the most important part of the instruction that shows what operation the instruction performs. There are 256 different values for the opcode that can be presented using a 256-dimensional vector.

ModRM and SIB are one byte each and follow the same structure, where the eight bits are divided into two, three, and three-bit parts. Each respective part can have four, eight, and eight possible values. Therefore, both ModRM and SIB can be presented using a 20-dimensional one-hot vector.

Finally, displacement and immediate are constant values, where the former is the operand address and the latter is the operand value. Displacement is almost always 32 bits and the operand is 64 bits. In total, we need a 96-dimensional vector to represent these two values.

Other than the opcode, the rest are optional and we need to indicate, which parts are present in the instruction. We use a 5-dimensional vector to show the presence of prefix bytes, ModRM, SIB, displacement, and immediate. In total the final vector size is $5+9+256+20+20+32+64=406$. After getting the final encoded vector for each instruction of a node, all the vectors are aggregated and stored for the next step.

In the second step, after preparing the initial encode vector, we train an autoencoder using only benign samples and try to minimize the reconstruction error. Then, the encoder is used to reduce the dimension of the rule-based encoded vector from 406 to 64.

\subsection{Graph Reduction}
Graph reduction is a crucial technique in graph-based machine learning, aimed at simplifying complex graphs by selectively removing nodes, edges, or subgraphs without significantly impacting the overall structure or performance of down-stream tasks. This process helps in reducing computational overhead, enhancing model interpretability, and mitigating overfitting by eliminating redundant or irrelevant components. Pruning strategies are particularly valuable in large-scale graph applications, where the sheer size and complexity of the data can hinder efficient analysis and model accuracy. In this section, we explore various graph pruning methodologies and their underlying principles.

\subsubsection{Leaf Prune (LP)}
Leaf pruning, our proposed graph reduction technique, which is tailored to the structure and topology of staticly generated CFGs and FCGs, is a graph reduction technique, where nodes that are identified as leaves—nodes with a degree less than or equal to 1—are removed from the graph. Unlike iterative methods, this approach typically involves a single step or a predefined condition to remove all leaf nodes at once. The primary objective of leaf pruning is to simplify the graph by eliminating nodes that contribute minimally to the overall connectivity, thereby retaining the more central structure of the graph.

In this method, the leaf nodes are first identified based on their degree. A node $v$ is considered a leaf if its degree $d(v)\leq1$. Once identified, these leaf nodes are removed from the graph along with their incident edges. This process is generally performed in a single operation, where all leaf nodes are pruned simultaneously, rather than iteratively checking and pruning new leaf nodes that may emerge.

Mathematically, let $G=(V, E)$ be a graph with a node set $V$ and an edge set $E$. The degree $d(v)$ of a node $v$ is defined as the number of edges incident to $v$. The set of leaf nodes $L$ can be described as:

\begin{equation}
    L=\left\{v \in V \mid d \left ( v \right ) \leq 1 \right\}
\end{equation}

The pruned graph $G' = (V', E')$ is then defined as:

\begin{equation}
    V' = V \setminus L
\end{equation}

This approach is particularly effective in scenarios, where the goal is to quickly reduce the graph's size by removing nodes that have minimal impact on its structure, such as in simplifying large networks for visualization or in pre-processing steps for more complex graph algorithms. Leaf pruning is suitable for network analysis, where understanding the core structure of the graph is more important than the peripheral, less connected nodes.

\subsubsection{Comp Prune (CP)}
Another graph reduction technique, which is proposed in this work, called comp prune. Comp Prune aims at simplifying a graph by removing a certain percentage of its smallest connected components. This method is particularly useful in filtering out less significant parts of a graph, such as noise or isolated clusters, allowing for a more focused analysis of the core structure of the graph topology.

The comp prune process begins by identifying all the connected components of the graph $G = (V, E)$. A connected component is defined as a maximal subgraph, where there exists a path between any two nodes. Once the components are identified, they are sorted based on their size, typically by the number of nodes within each component.

Let $ C_1, C_2, \dots, C_n $ represent the sorted connected components of $G$, with $|C_i|$ denoting the size of component $C_i$ (i.e., the number of nodes). The components are sorted such that $ |C_1| \leq |C_2| \leq \dots \leq |C_n| $. The pruning is performed by removing the smallest $u \times n $ components, where $u$ is a user-defined pruning parameter representing the percentage of components to be removed.

Comp prune is an effective strategy for graph simplification, especially in scenarios where the focus is on the more densely connected and significant parts of the network, such as in social network analysis, biological network studies, and other applications, where large graphs need to be reduced without losing critical structural information

\subsubsection{K-core}
K-core decomposition is a state-of-the-art technique in graph theory used to analyze the structure of a graph by iteratively removing nodes with degrees less than $k$~\cite{hashemi2024comprehensive}. A $k$-core of a graph is a maximal subgraph, in which each node has at least $k$ neighbors. This technique is valuable for identifying tightly connected communities within a graph and has applications in various domains, including social network analysis, biology, and data mining.

The process of k-core decomposition involves progressively pruning nodes from the graph. Initially, all nodes with a degree less than $k$ are removed. This reduction may cause other nodes to drop below the degree threshold $k$, leading to further pruning. The process continues iteratively until no more nodes can be removed. The resulting subgraph is the $k$-core of the original graph, and it represents the most cohesive structure within the graph, where all remaining nodes meet the degree condition.

Mathematically, let $ G = (V, E) $ be a graph with a node set $V$ and an edge set $E$. The degree $d(v)$ of a node $v$ is the number of edges incident to $v$. The k-core of $G$, denoted as $ G_k = (V_k, E_k) $, is defined as:

\begin{equation}
    G_{k}=\left\{ v \in V \mid d_{G_{k}} \left ( v \right ) \geq k \right\}
\end{equation}

where $ d_{G_k}(v) $ represents the degree of node $v$ within the subgraph $G_k$. The k-core decomposition can be computed efficiently using a simple algorithm that operates in linear time relative to the number of edges in the graph.

The k-core decomposition is particularly useful for uncovering the ``core" structure of a network, highlighting the most interconnected and resilient nodes. This property makes it a powerful tool in various applications, such as identifying influential nodes in social networks, detecting protein complexes in biological networks, and simplifying complex graphs for visualization and analysis. The iterative nature of the decomposition ensures that the resulting k-core subgraph is maximally connected, preserving the essential connectivity of the original graph~\cite{hashemi2024comprehensive}.

\subsubsection{Walk Index Sparsification}
Walk Index Sparsification (WIS) is a state-of-the-art graph sparsification technique designed to reduce the number of edges in a graph, while preserving its critical structural properties, particularly those that influence the graph's ability to model interactions effectively~\cite{razin2023ability}. WIS focuses on maintaining the integrity of the graph's walk index, which measures the number of walks (or paths) of a given length between pairs of nodes. This index is crucial in applications such as GNNs, where the capacity to model node interactions is essential.

The walk index of a graph is closely tied to its ability to propagate information across its structure. The primary objective of WIS is to prune edges that have minimal impact on this information propagation capability. The technique evaluates edges based on the degrees of the nodes they connect, using criteria such as \texttt{deg\_min} (the minimum degree of the two nodes) and \texttt{deg\_max} (the maximum degree of the two nodes). By focusing on edges with the lowest \texttt{deg\_min}, WIS ensures that the most critical connections within the graph are preserved, while less significant edges are removed.

Mathematically, the process begins by calculating the degrees of all nodes in the graph. For each edge $ e = (u, v) $, where $u$ and $v$ are nodes on the two ends of $e$, the values of \texttt{deg\_min}(e) and \texttt{deg\_max}(e) are defined as follows:

\begin{equation}
    \texttt{deg\_min}(e) = \min(\text{deg}(u), \text{deg}(v))
\end{equation}

\begin{equation}
    \texttt{deg\_max}(e) = \max(\text{deg}(u), \text{deg}(v))
\end{equation}

Edges are then ranked primarily by \texttt{deg\_min}, with \texttt{deg\_max} and lexicographic order used to break ties. The edge with the smallest \texttt{deg\_min} is considered the least crucial for maintaining the graph's walk index and is therefore the first candidate for removal. This process is repeated iteratively, with edges being removed in order of their rank until the desired level of sparsification is achieved.

The effectiveness of WIS lies in its ability to preserve the graph's walk index, even as edges are pruned. Doing so maintains the graph's capability to model interactions, which is essential in tasks such as node classification, link prediction, and other GNN-based applications. The methodology provides a balanced approach to graph sparsification, ensuring that the reduction in complexity does not compromise the graph's functional integrity~\cite{razin2023abilitygraphneuralnetworks}.

\subsection{Graph Classification}
\begin{figure}
    \centering
    \includegraphics[width=0.8\linewidth]{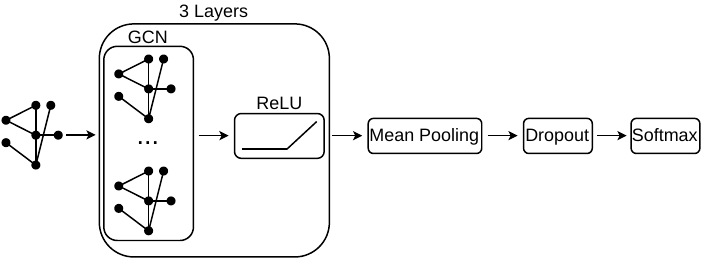}
    \caption{Graph classification model architecture.}
    \label{fig:gcn}
\end{figure}

We use an architecture based on a GCN for the graph classification module. Figure~\ref{fig:gcn} shows the architecture of the classification model. It consists of three layers of graph convolution network and ReLU followed by a mean pooling and a dropout layer. At the end, there is a softmax layer to do the classification and output the class probabilities.

\subsection{Explainability}
As discussed before there has been a huge amount of research regarding the explainability of graph classification models, and several state-of-the-art techniques are proposed focusing on providing factual and counterfactual explanations for these models. Among those, GNNExplainer~\cite{ying2019gnnexplainer} is a state-of-the-art and most cited techniques that provide important sub-graphs with regard to the model decision as the explanation. The main intuition is to find a sub-graph that maximizes the mutual information with the model's prediction. This can be formulated using the following equation:

\begin{equation}
    \underset{G_{s}}{max}MI(Y,(G_{s},X))=H(Y)-H(Y|G=G_{s}, X)
\end{equation}

where $Y$ is the prediction distribution of the model, $X$ the node feature vector and $G_{s}$ is the important sub-graph for the model's prediction. Since for a trained GNN, $H(Y)$ is constant, to maximize the above equation it is only required to minimize the conditional entropy. The conditional entropy can be replaced with a cross-entropy loss between the true class label and the model prediction to reach the following equation, which can be optimized using gradient descent:

\begin{equation}
    \underset{M}{min}-\sum_{c=1}^{C}\mathds{1}[y=c]logP_{\Phi}(Y=y|G=A_{c}\odot\sigma(M),X)
\end{equation}

where $c$ is the class labels, $y$ is the predicted label, $A_{c}$ is the adjacency matrix of the input graph, $M$ is the learnable mask that gives a weight to each edge of the input graph based on their importance for the model's prediction, $\sigma$ is the sigmoid function to map the weight to $[0,1]$, and $X$ the node feature vector. Finally, a threshold will be used to remove the edges with low weight, to reach the explanation sub-graph.
\section{Results and Analysis}\label{sec:result}
In our work, we construct a comprehensive static pipeline for the analysis and classification of malicious and benign binary files. Towards this pipeline, we use several programming languages and associated libraries. Most notably angr, PyTorch~\cite{paszke2017automatic}, PyTorch Geometric~\cite{DBLP:journals/corr/abs-1903-02428}, and Networkx~\cite{paper:hagberg:2008}.

Numerous technical challenges are present in every stage of our pipeline. However, almost all major challenges can be traced to the large size of sample graphs. Particularly stemming from malicious graphs in the BODMAS dataset. Such large graphs generate extreme costs in terms of both time and space. Most notably, apart from significantly impacting more heavyweight algorithms for training, pruning, and explanation the largest challenge associated with such large graphs is the serializing and deserializing operations needed to access each individual graph before and after any processing or transformation operation is needed. When such operations are repeated over a large search space with multiple epochs worth of access these costs in both time and space climb rapidly and form the largest observable bottleneck in the pipeline.

\subsection{Datasets}
In our experimental setup, we use three main datasets including BODMAS, DikeDataset, and PMML.

The Blue Hexagon Open Dataset for Malware AnalysiS (BODMAS) dataset (2019-2020)~\cite{yang2021bodmas} contains 134,435 PE samples, including 57,393 malware and 77,142 benign samples. These samples were randomly selected from an unnamed company's internal malware database. BODMAS focuses exclusively on binary samples for malicious files and supports detection or classification into 581 distinct malware families. The dataset is timestamped and represented as a 2381-feature vector, consistent with other datasets such as Ember and SOREL20M, facilitating dataset extension.

The DikeDataset (2021)~\cite{dikedataset} comprises 982 benign PE files, 100 benign OLE files, 8,970 malicious PE files, and 1,871 malicious OLE files. These samples were sourced from various repositories, including the Malware Detection PE-Based Analysis Using Deep Learning Algorithm Dataset, MalwareBazaar, and DuckDuckGo. Notably, DikeDataset includes benign file samples, which are valuable for the classification and detection of malware using machine learning models. The samples are labeled on a scale from 0 to 1 based on specific malware family labels.

The PE Malware Machine Learning Dataset (PMML), provided by Practical Security Analytics, includes 201,549 labeled portable executable samples, with 86,812 legitimate and 114,737 malicious files~\cite{practicalsecurity2024pe}. The dataset, spanning 117GB uncompressed, offers detailed metadata, including file hashes, antivirus scan results, file types, and entropy. While extensive, PMML predominantly contains easily accessible malware, leading to a prevalence of similar malware families. Although it includes samples from Advanced Persistent Threat (APT) actors, these are outnumbered by generic adware, spyware, and ransomware samples. Additionally, the legitimate files are primarily sourced from various versions of Windows, especially Microsoft-produced software, potentially introducing bias. PMML is designed to support comprehensive malware analysis and enhance security and AI research by providing raw labeled data instead of pre-extracted features.

While many benign and malicious samples exist across the aforementioned datasets we restrict ourselves to file samples of type PE, the focus of our work, and the architecture of type x86 to bring our analysis and methods in line with others~\cite{gao2022malware}. Specifically, for each sample we compute the output string of the standard `file' command, and, then, test membership against sets of both whitelist and blacklist regular expressions. We found that even though some datasets contained both malicious and benign samples only one class contained the required sample types. In total we extract 1188 and 1125 malicious samples from the BODMAS and PMML datasets, respectively, as well as 520 benign samples from the DikeDataset. A copy of our dataset is available \href{https://www.unb.ca/cic/datasets/sgg-dataset-2024.html}{online}.

\begin{figure*}
    \captionsetup[subfigure]{justification=centering}
    \centering
    \begin{subfigure}[b]{0.24\textwidth}
        \centering
        \includegraphics[width=\textwidth]{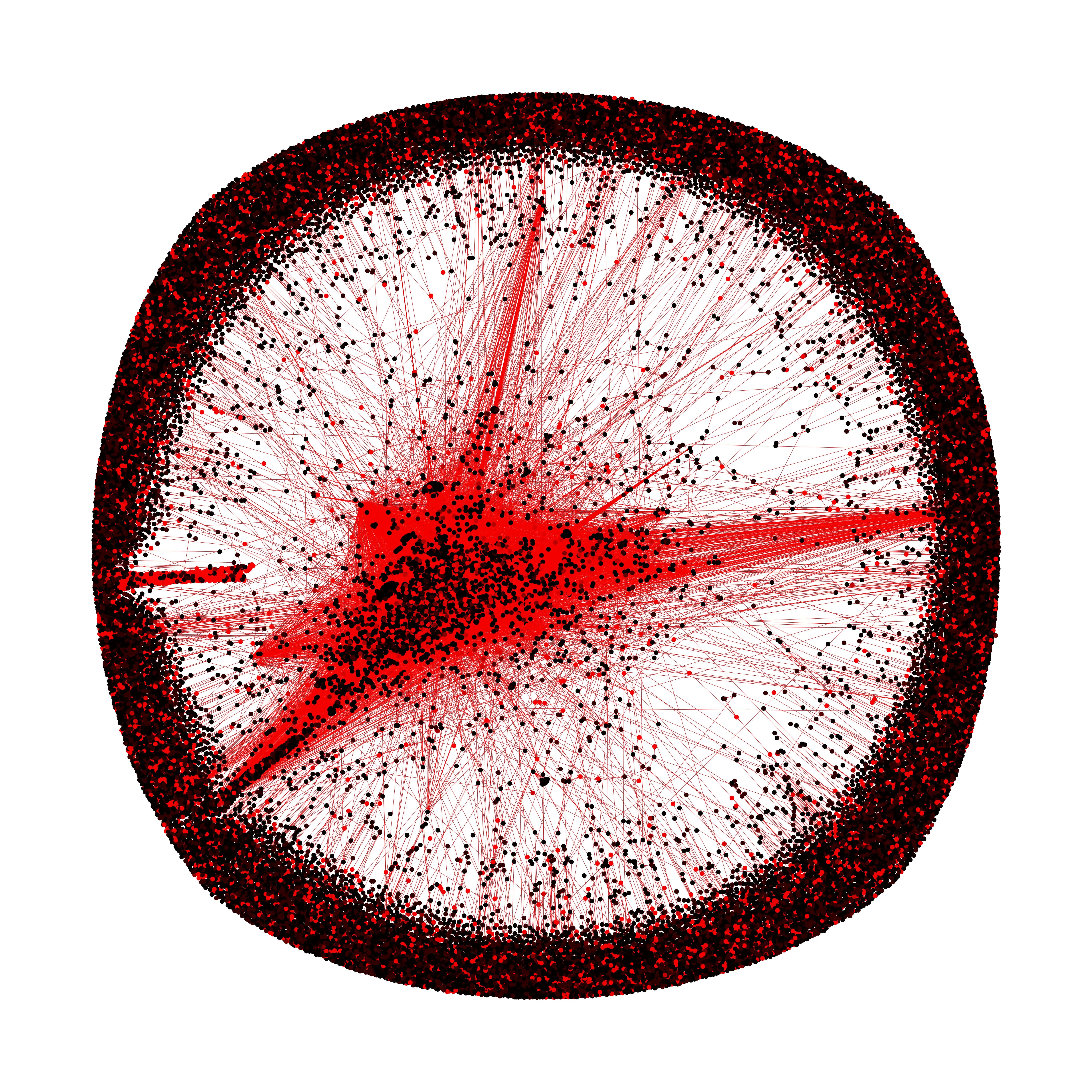}
        \caption{Malicious - FNE (FCG) - No Prune (NP)\\(N:34,278 - E:22,398)}
    \end{subfigure}
    \begin{subfigure}[b]{0.24\textwidth}
         \centering
         \includegraphics[width=\textwidth]{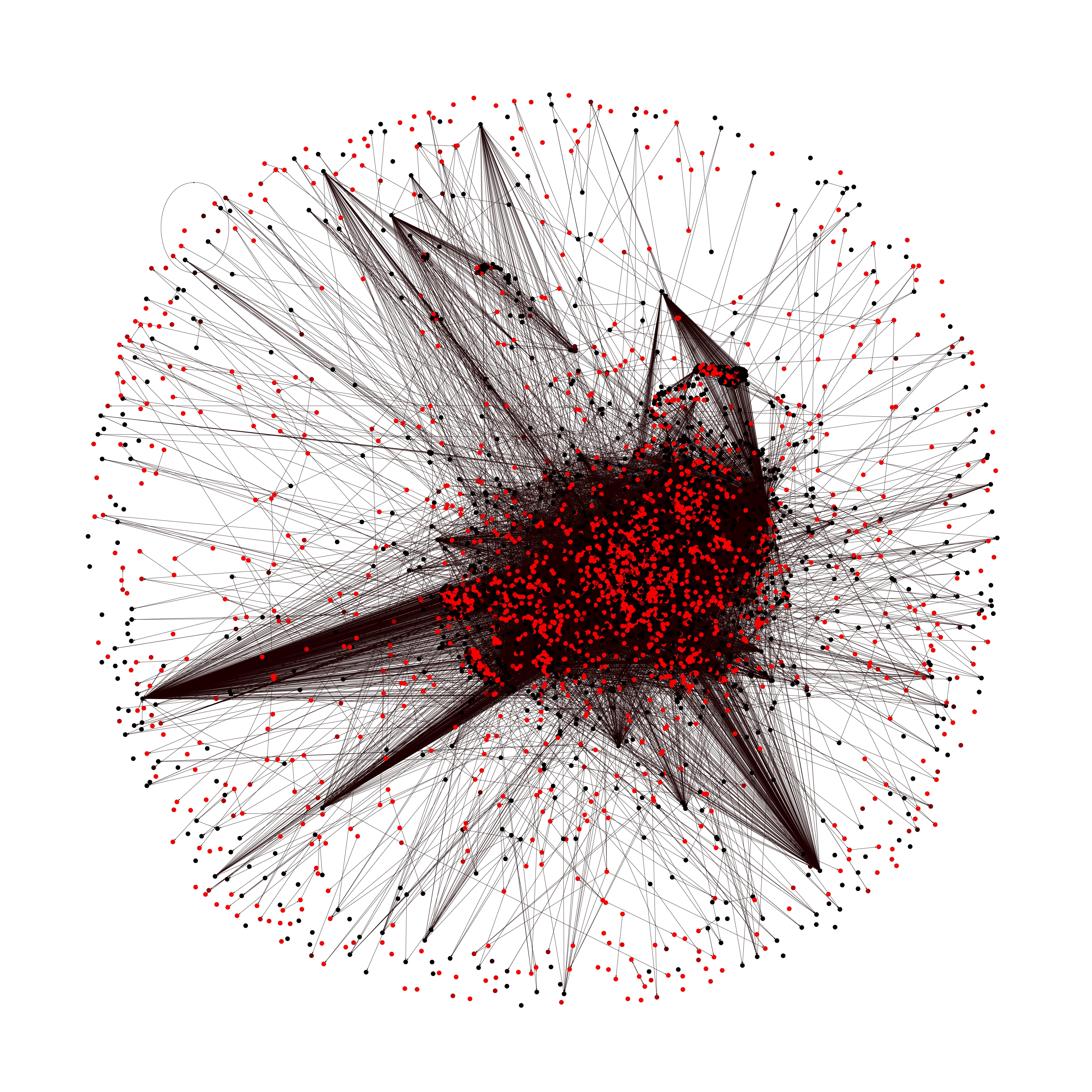}
         \caption{Malicious - FNE (FCG) - Leaf Prune (LP)\\(N:4890 - E:18,989)}
    \end{subfigure}
    \begin{subfigure}[b]{0.24\textwidth}
        \centering
        \includegraphics[width=\textwidth]{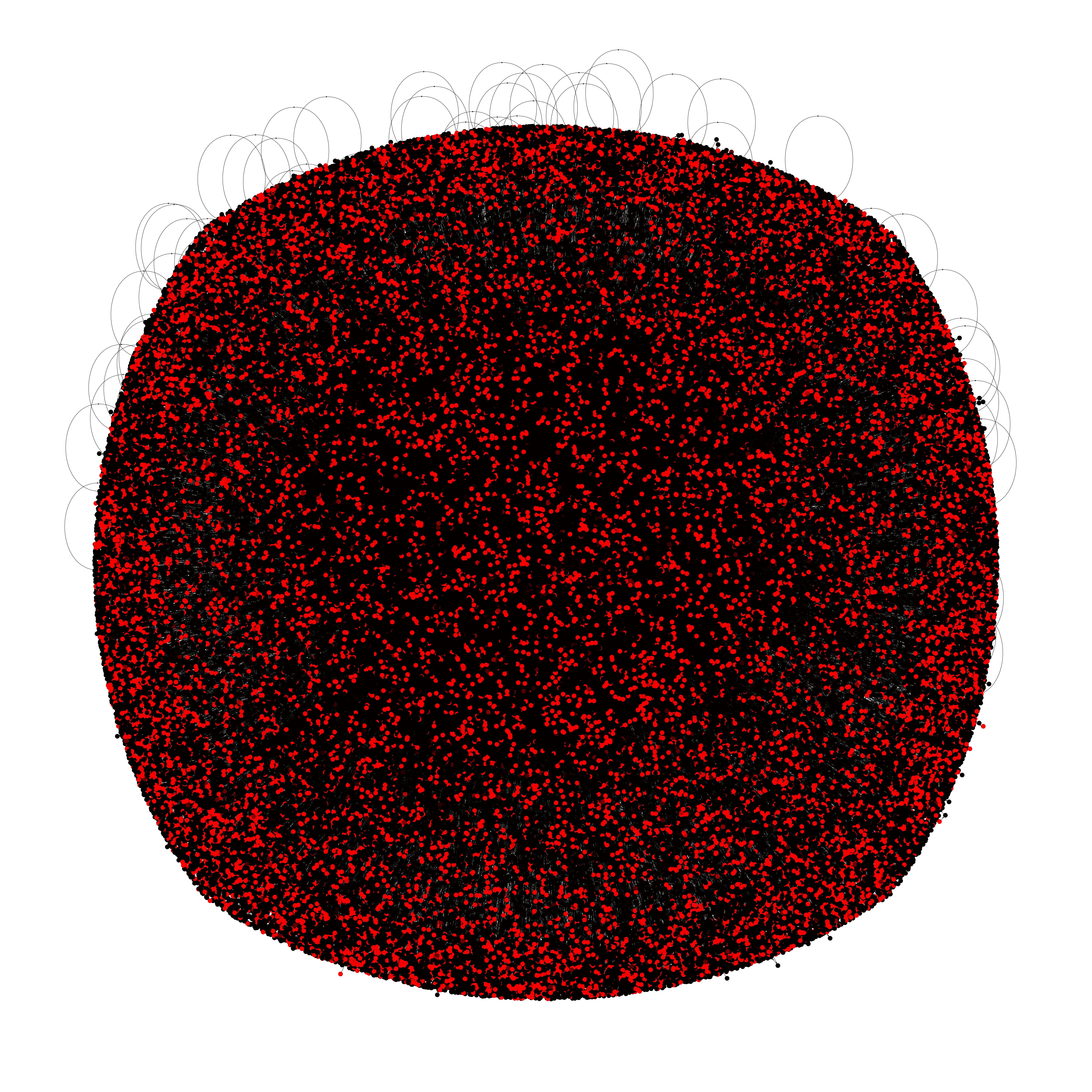}
        \caption{Malicious - AE (CFG) - NP\\(N:169,513 - E:265,538)}
    \end{subfigure}
    \begin{subfigure}[b]{0.24\textwidth}
         \centering
         \includegraphics[width=\textwidth]{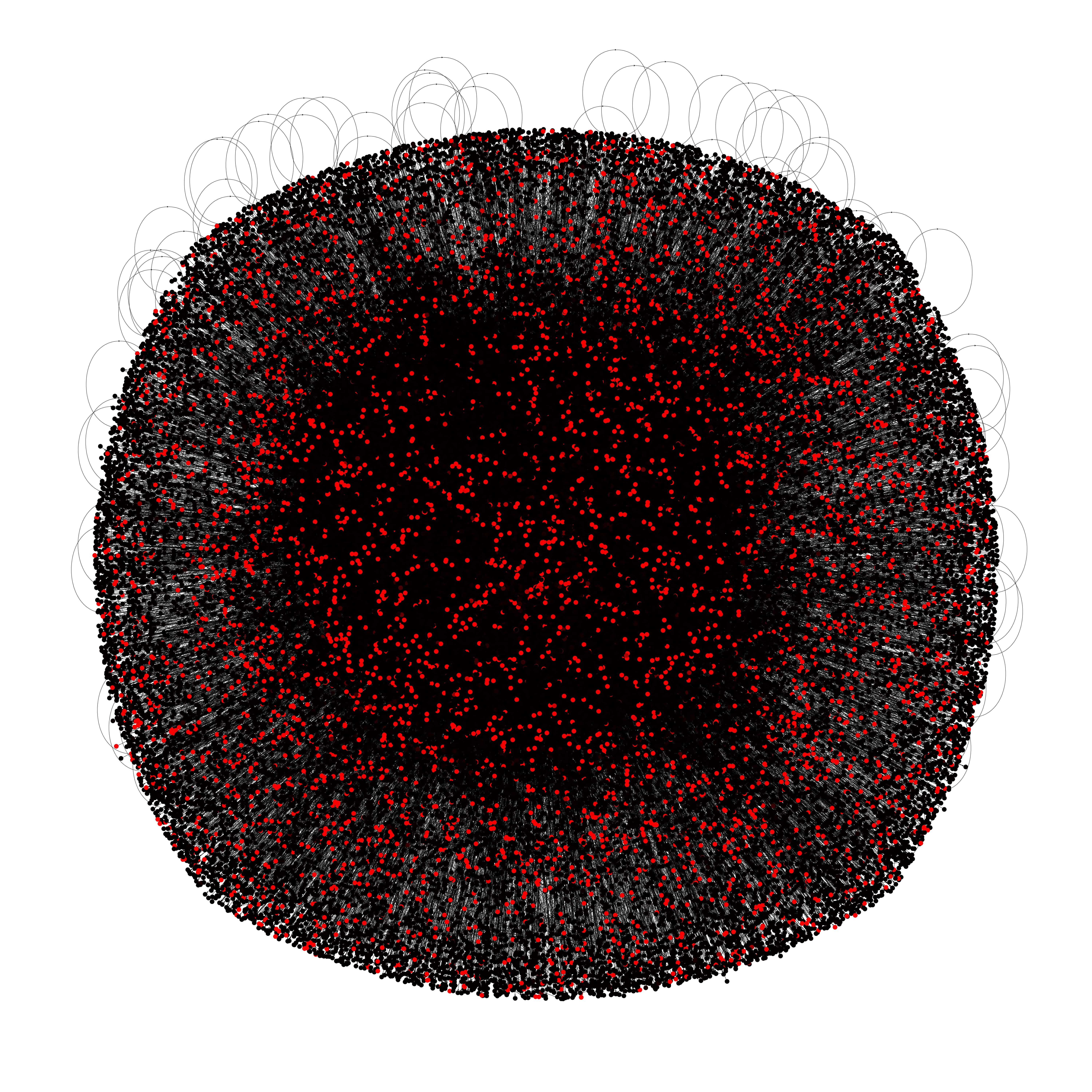}
         \caption{Malicious - AE (CFG) - LP\\(N:127,066 - E:242,973)}
    \end{subfigure}
    \\
    \begin{subfigure}[b]{0.24\textwidth}
        \centering
        \includegraphics[width=\textwidth]{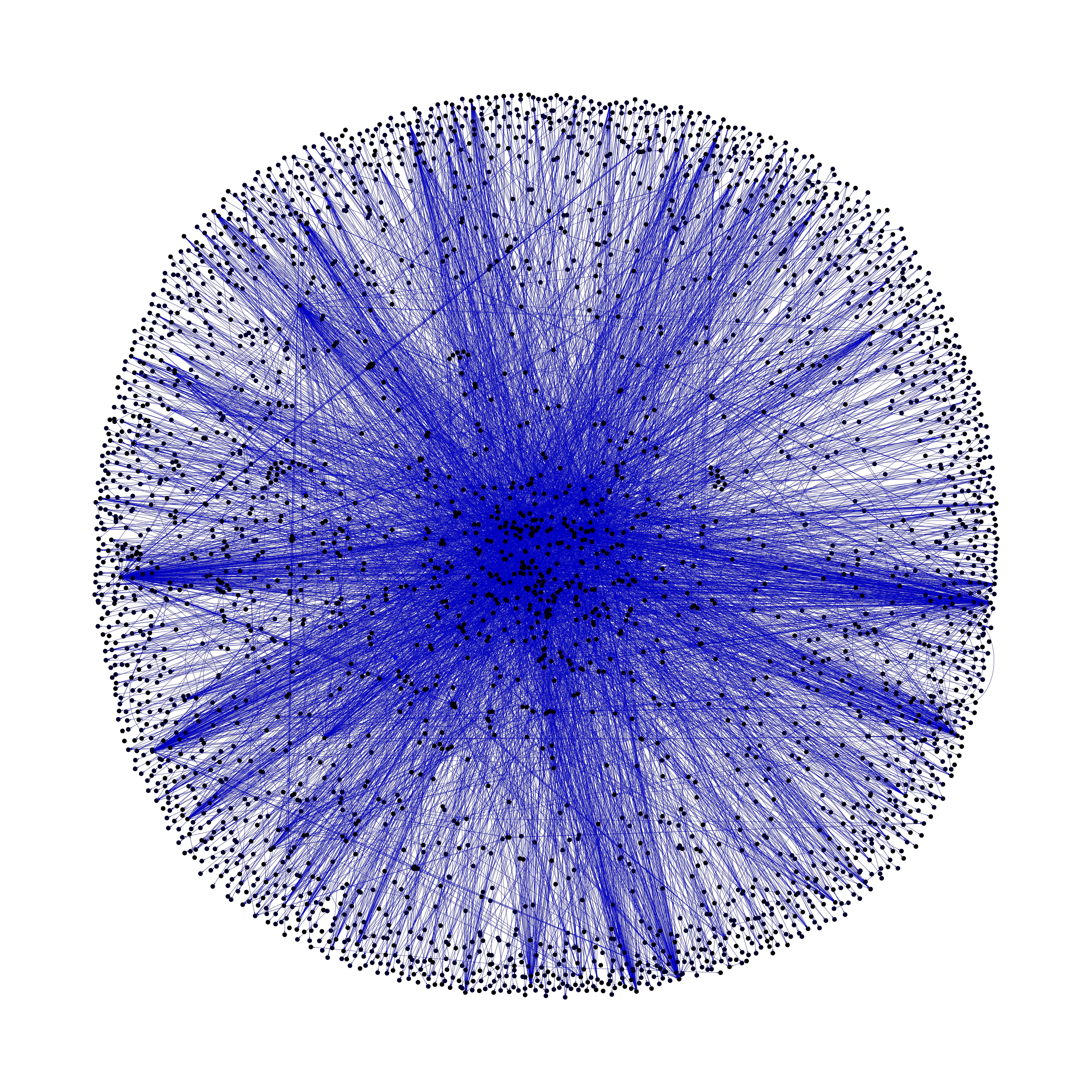}
        \caption{Benign - FNE (FCG) - NP\\(N:3,243 - E:6,506)}
    \end{subfigure}
    \begin{subfigure}[b]{0.24\textwidth}
         \centering         
         \includegraphics[width=\textwidth]{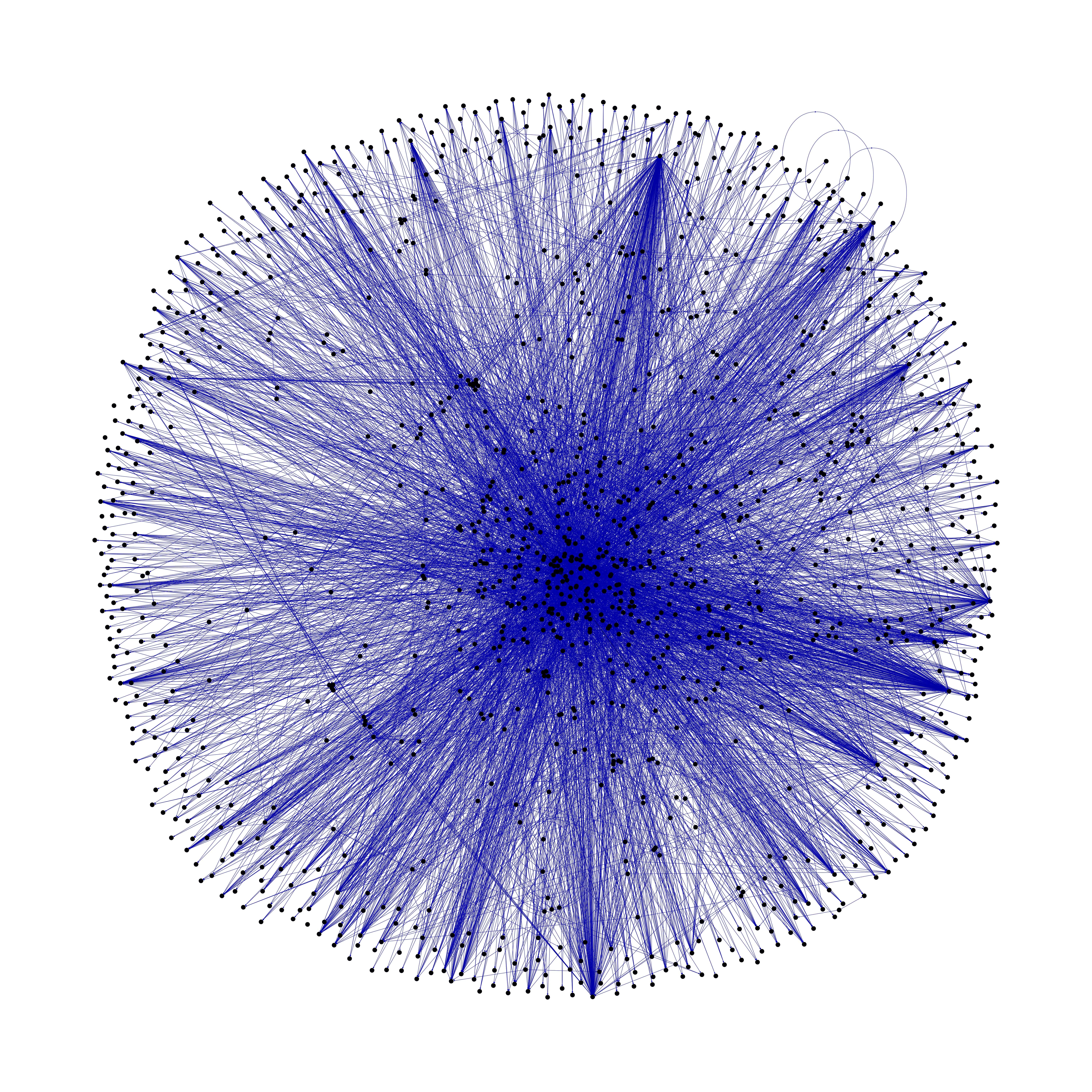}
         \caption{Benign - FNE (FCG) - LP\\(N:1,250 - E:5,213)}
    \end{subfigure}
    \begin{subfigure}[b]{0.24\textwidth}
        \centering
        \includegraphics[width=\textwidth]{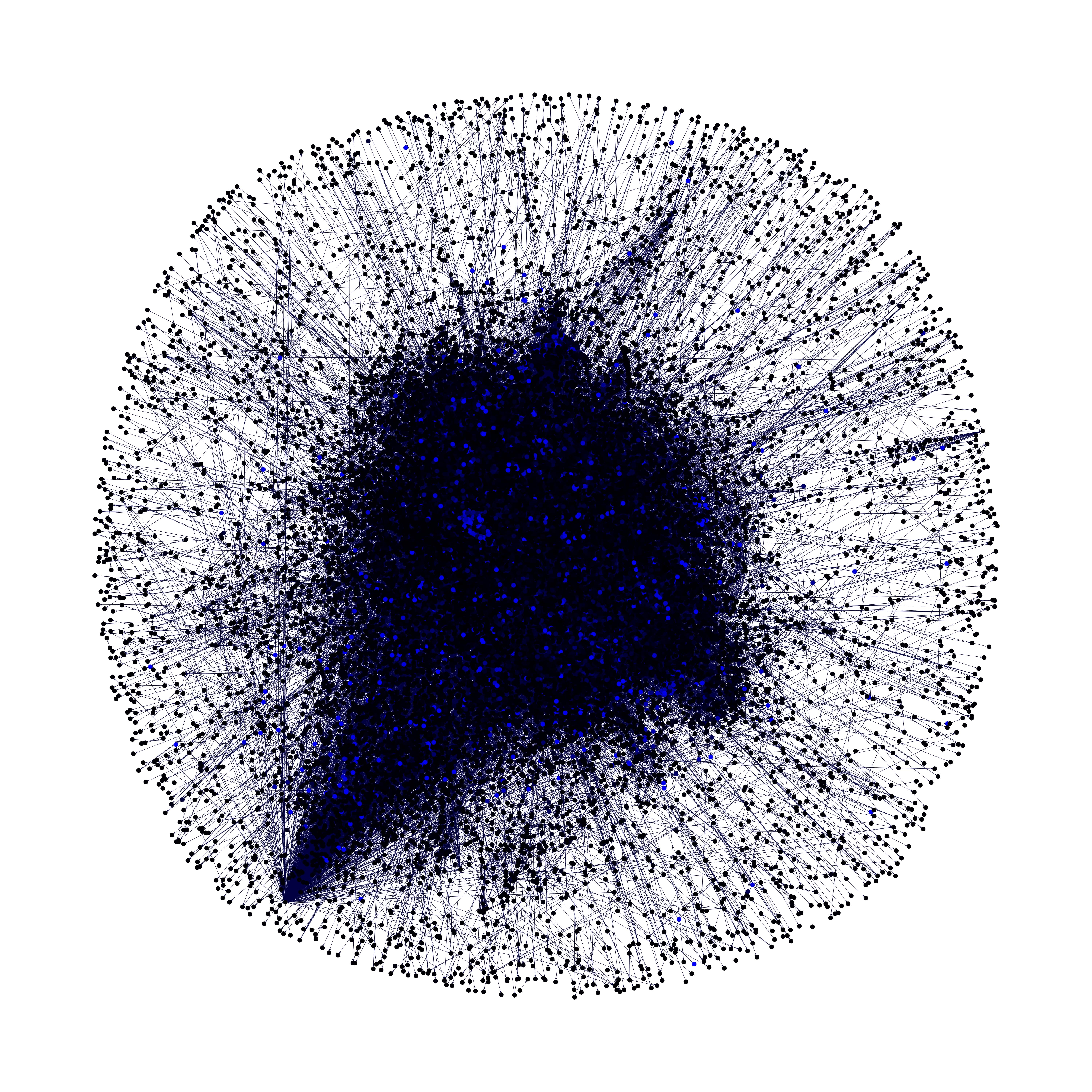}
        \caption{Benign - AE (CFG) - NP\\(N:28,930 - E:62,303)}
    \end{subfigure}
    \begin{subfigure}[b]{0.24\textwidth}
         \centering
         \includegraphics[width=\textwidth]{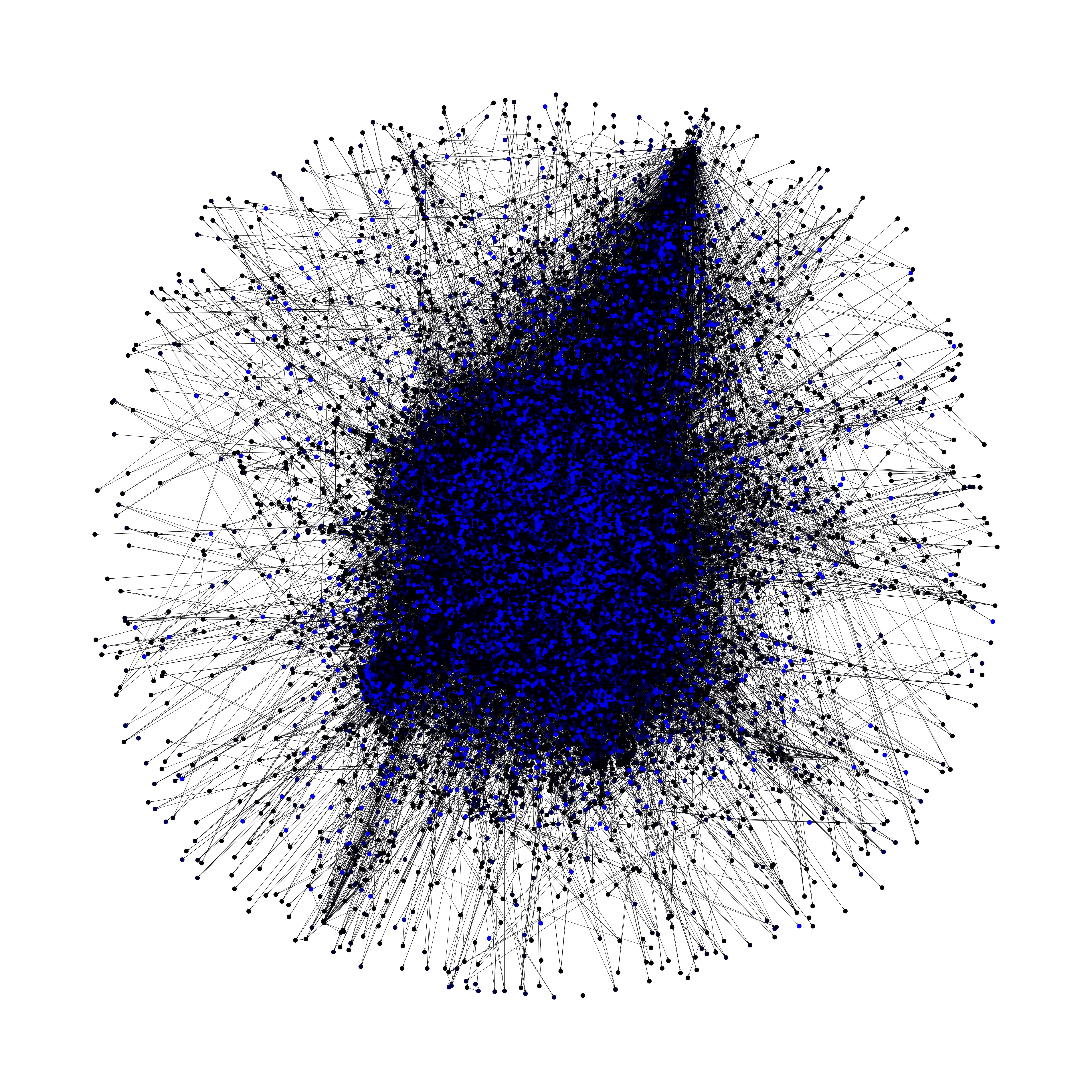}
         \caption{Benign - AE (CFG) - LP\\(N:27,224 - E:61,281)}
     \end{subfigure}
     \caption{Visualization of example malicious and benign samples. Each panel shows a generated graph through an embedding module with/without pruning. NP and LP stand for Leaf Prune and No Prune, respectively.}
     \label{fig:vis}
\end{figure*}

\subsection{Visualization}
To provide a more clear understanding of the samples and graphs used in the evaluation, in Figure~\ref{fig:vis}, we visualize one malicious and one benign sample, with and without pruning generated through two proposed embedding methods. As it is clear in this figure the number of nodes and edges are a lot more for FNE graphs than the AE graphs. This reflects the considerable size difference between FCGs and CFGs, which is due to the information that they provide in each node. FCG includes only function calls, while CFG records all the basic blocks, which may include several conditions, loops, and transitions. When comparing the size of the graphs with and without pruning again there is a considerable difference between FNE and AE graphs. It seems that leaf prune is more effective on FCG and reduces its size significantly. Since leaf prune only prunes nodes with degree less than or equal to one it shows that FCGs have so many leaf and island nodes.

\subsection{Graph-based Analysis}
\begin{figure*}[!b]
    \centering
    \includegraphics[width=.7\textwidth]{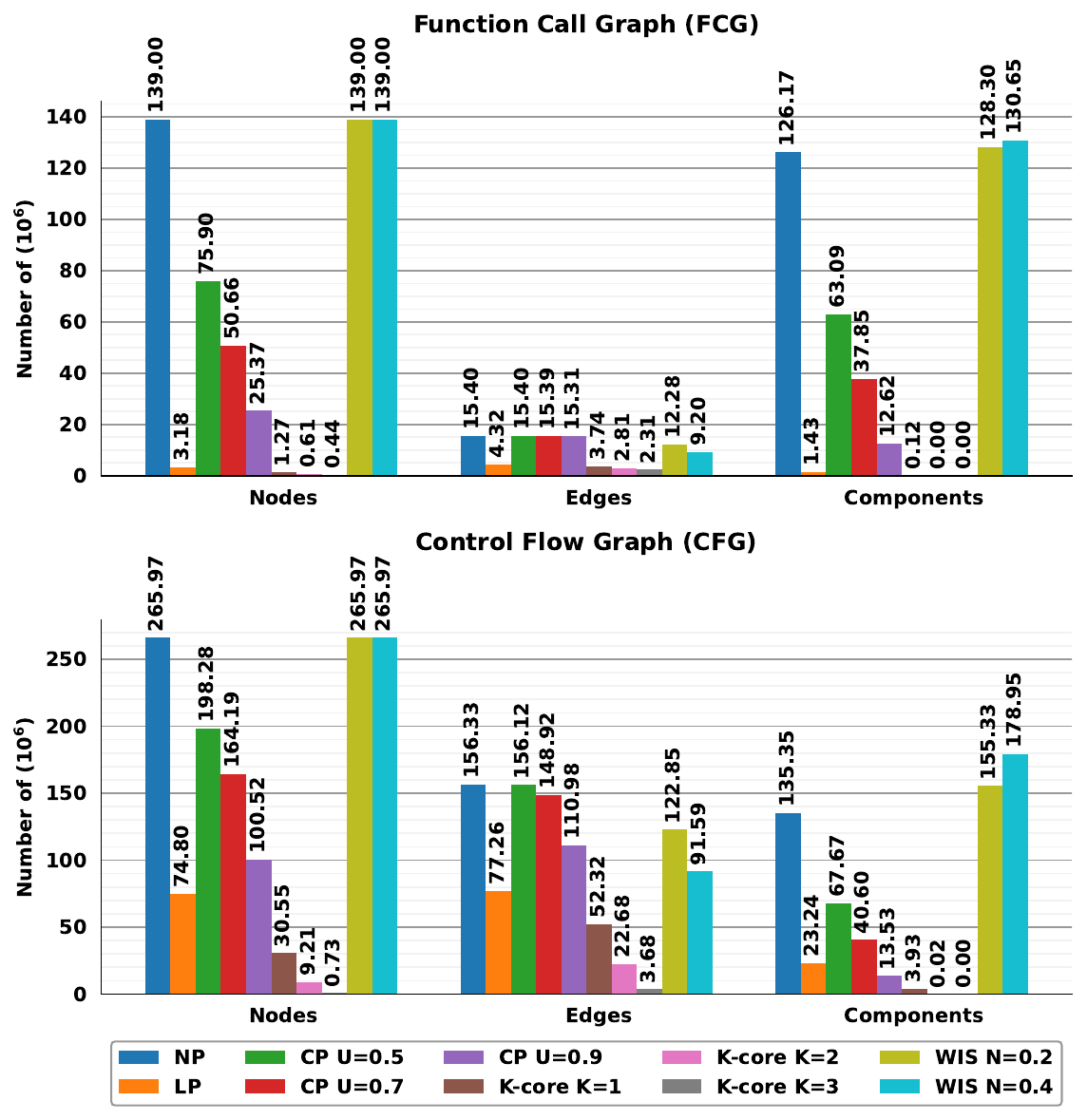}
    \caption{Comparing number of nodes, edges, and components of FCG (top panel) and CFG (bottom panel) generated through each graph reduction technique, with no pruning as baseline.}
    \label{fig:graph_prune}
\end{figure*}

The first part of the analysis is focused on the performance of the graph reduction methods with regard to the number of nodes, edges, and connected components. The goal here is to reduce the size of the graphs to make the detection process more efficient and accurate. As mentioned before each of FNE and AE are compatible with a different graph structure, therefore both FCG and CFG are generated in the graph generation step. The size and details of these two types of graphs are different, therefore, two separate charts are used to present the results and compare the number of nodes, edges, and components of pruned graphs by each reduction method and the original graphs. Figure~\ref{fig:graph_prune} compares the number of nodes, edges, and components for both FCG and CFG with different reduction methods including No Pruning (NP), Leaf Prune (LP), Comp Prune (CP), K-core, and WIS. For Comp Prune, we did the experiments with $u$ equal to 0.5, 0.7, and 0.9, which is the percentage of the components to be removed. K-core was done with $k$ equal to 1, 2, and 3, and for WIS, 20 and 40 percent of the top-ranked edges were removed, respectively.

For FCGs, NP represents the baseline configuration with 139.00 million nodes, 15.4 million edges, and 126.17 million components. The leaf prune technique yields significant changes to the graph structure, indicating its effectiveness in reducing graph size. Comp prune, on the other hand, demonstrates a progressive reduction in nodes and edges as the threshold $U$ increases from 0.5 to 0.9, with $U=0.9$ achieving the most significant reduction. However, the graph size is still bigger compared to leaf prune.  Similarly, K-core shows a sharp decrease in graph complexity, with higher $K$ values ($K=3$) yielding the most substantial reductions (almost zero), which may result in losing valuable information and is not desirable for the detection step. The WIS technique exhibits a moderate reduction in graph size since it only prunes edges and keeps the nodes unchanged, with $N=0.4$ outperforming $N=0.2$.

For CFGs, NP again establishes a large baseline graph with 265.97 million nodes, 156.33 million edges, and 135.35 million components. In contrast to FCGs, leaf prune has a lesser effect, while resulting in a significant reduction in nodes and edges. Similar to the FCG results, comp prune shows a strong correlation between increasing threshold $U$ and decreasing graph size, with $U=0.9$ producing the most compact graph. K-core behaves similarly to its performance on FCGs, with higher $K$ values leading to substantial reductions in nodes and edges. WIS also reduces graph complexity, with $N=0.4$ producing a more condensed graph than $N=0.2$, though its overall reduction remains less aggressive compared to CP and K-core.

In summary, the results demonstrate that the choice of graph reduction technique significantly impacts the size and complexity of both pruned FCGs and CFGs. Techniques such as comp prune and K-core are particularly effective with higher values for their hyper-parameters, which shows that they are dependent on the parameter selection. While WIS provides moderate reductions, it is less effective than comp prune and K-core in terms of graph size and more resource-intensive compared to leaf prune. Finally, leaf prune seems to be the most stable technique which reduces the size of both types of graphs considerably while working fast and efficiently. These findings highlight the importance of selecting the appropriate reduction technique based on the specific characteristics of the graph and the desired level of reduction.

\subsection{Detection-based Analysis}
\begin{figure}
    \centering
    \begin{subfigure}[b]{0.48\textwidth}
        \centering
        \includegraphics[width=\textwidth]{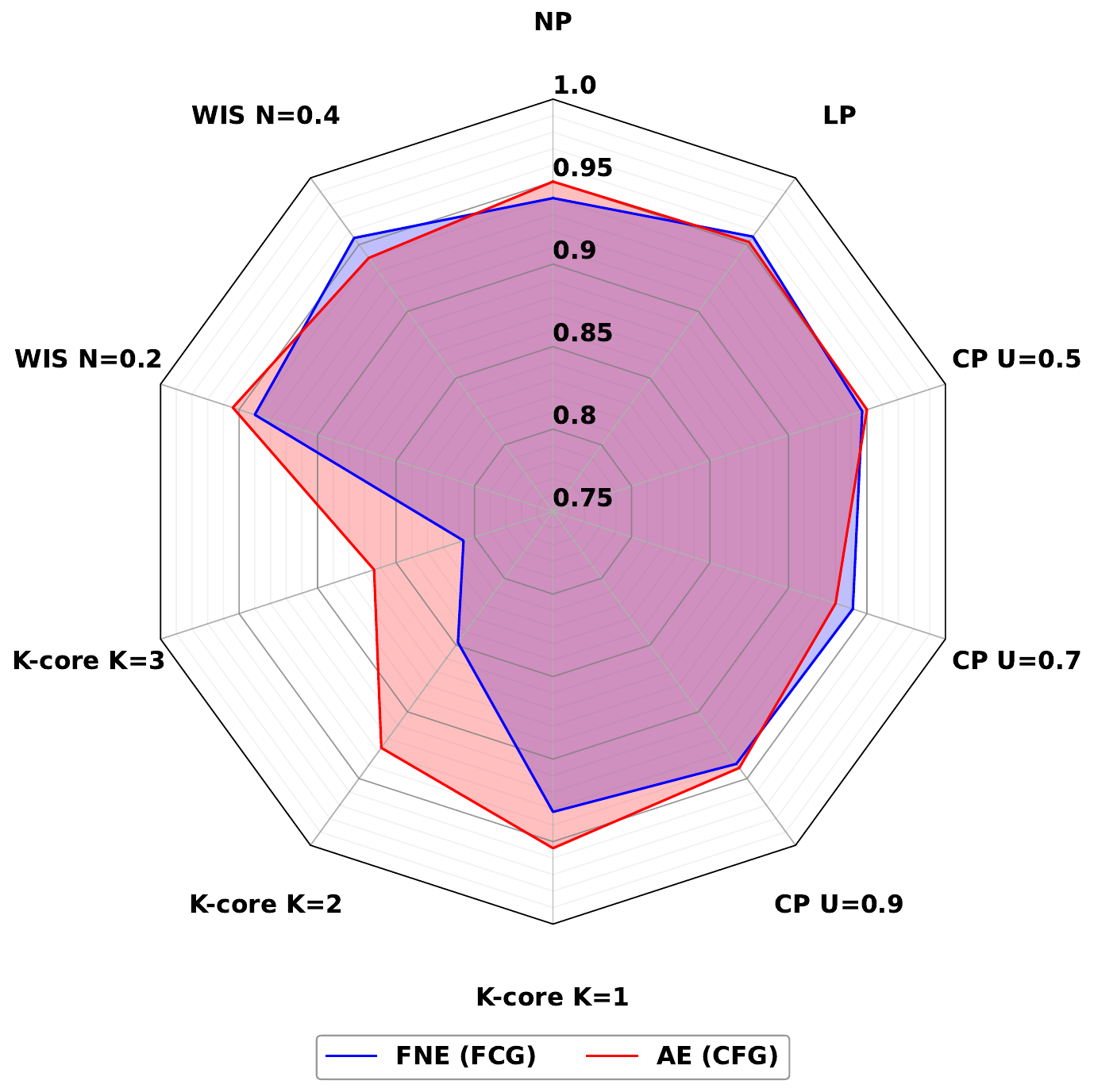}
        \caption{Accuracy}
     \end{subfigure}
     \begin{subfigure}[b]{0.48\textwidth}
         \centering
         \includegraphics[width=\textwidth]{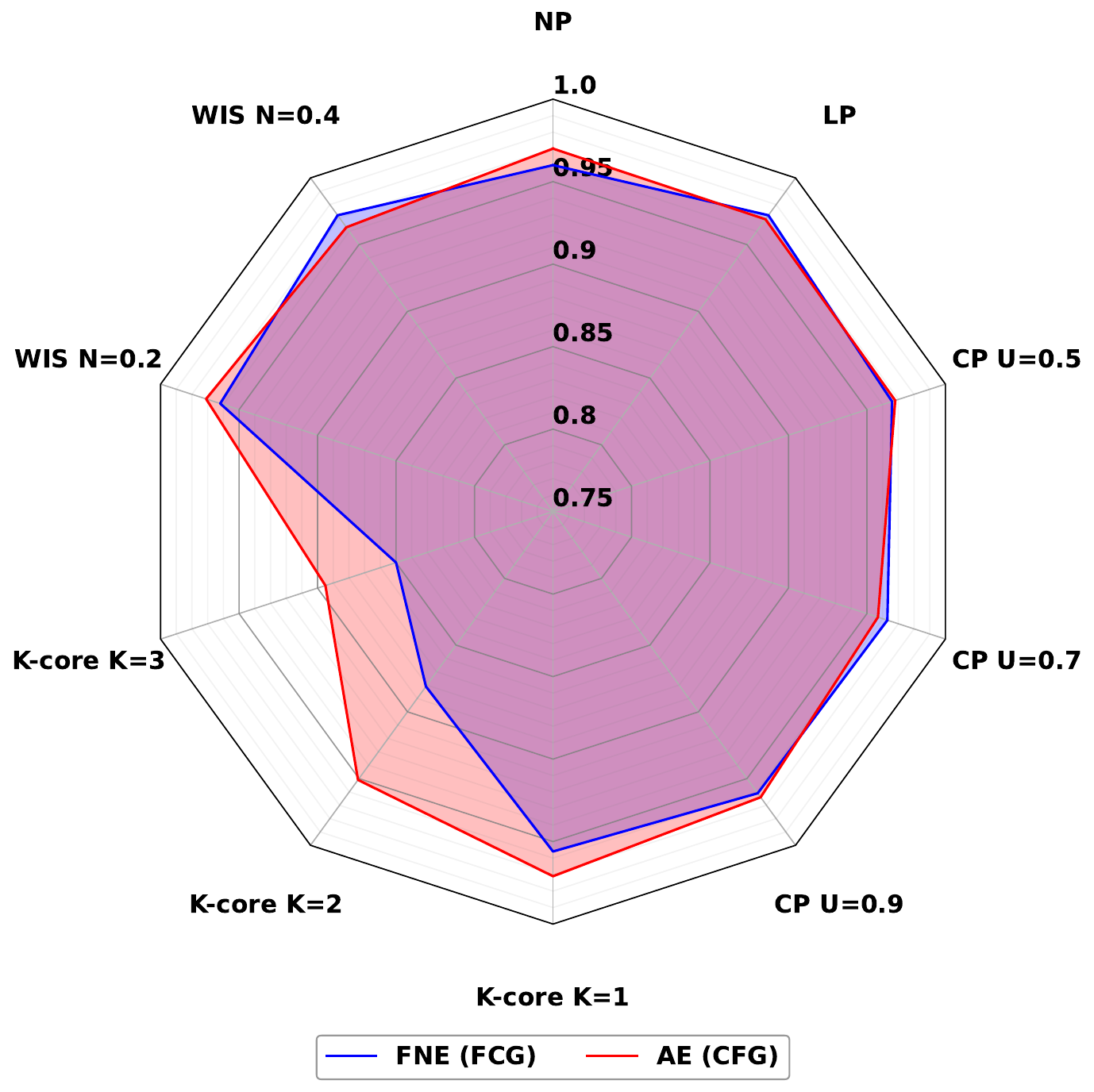}
         \caption{F1-score}
     \end{subfigure}
     \caption{Classification accuracy and F1-score obtained through each embedding technique along with each reduction technique and no reduction.}
     \label{fig:accf1}
\end{figure}

For the graph classification, we use a GCN-based architecture on top of each node embedding and graph reduction techniques. In Figure~\ref{fig:accf1}, we utilize a radar chart to compare the detection accuracy and F1-score of the two node embedding techniques more clearly. In general, it is clear that AE is working slightly better than FNE, and as expected it has better accuracy and F1-score due to the reason that it includes more detailed features of each node in the detection process. Without any pruning, the AE method has better performance, and even with different reduction techniques it still works better than FNE or very close to it except in the case of the comp prune with $u=0.7$ and WIS with $N=0.4$.

\begin{figure}
    \centering
    \includegraphics[width=.6\linewidth]{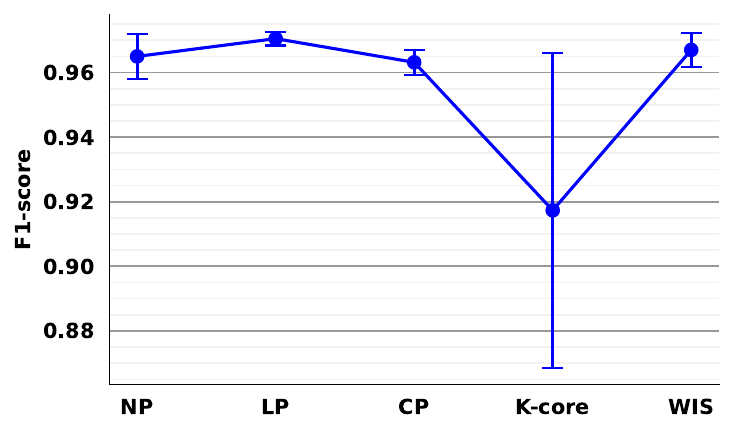}
    \caption{Average and standard deviation of F1-scores obtained through each graph reduction technique w.r.t. different hyper-parameters and node embeddings.}
    \label{fig:ave-std}
\end{figure}

To have a better comparison between the evaluated reduction methods, Figure~\ref{fig:ave-std} presents the average and standard deviation of F1-score of the evaluated techniques obtained through different node embedding. As it is evident leaf prune has better performance compared to other reduction methods. The details of the classification metrics are presented in Table~\ref{tab:ave_std}. Comparing the other metrics including accuracy, precision, and recall shows that leaf prune is the best reduction technique in all settings, while reducing the size of graphs significantly and making the learning and explainability process faster and less resource intensive. Bold entries in Table\ref{tab:ave_std} indicate highest average and lowest standard deviation achieved through each graph reduction technique.

\begin{table}
    \setlength{\tabcolsep}{2pt}
    \centering
    \caption{Average and standard deviation of classification metrics obtained through each graph reduction technique w.r.t. different hyper-parameters and node embeddings.}
    \scalebox{.8}{%
    \begin{tabular}{lcccc}
        \toprule
        \textbf{Pruning Method} & \textbf{Accuracy} & \textbf{F1-score} & \textbf{Precision} & \textbf{Recall} \\
        \midrule
        No Prune & 0.945$\pm$0.007 & 0.965$\pm$0.007 & 0.96$\pm$0.014 & 0.97$\pm$0.028\\
        \midrule
        Leaf Prune & \textbf{0.954$\pm$0.003} & \textbf{0.971$\pm$0.002} & \textbf{0.967$\pm$0.003} & 0.975$\pm$\textbf{0.001} \\
        \midrule
        Comp Prune & 0.942$\pm$0.007 & 0.963$\pm$0.004 & 0.956$\pm$0.018 & 0.971$\pm$0.015 \\
        \midrule
        K-core & 0.889$\pm$0.058 & 0.917$\pm$0.049 & 0.924$\pm$0.05 & 0.911$\pm$0.051 \\
        \midrule
        WIS & 0.947$\pm$0.008 & 0.967$\pm$0.005 & 0.953$\pm$0.01 & \textbf{0.98}$\pm$0.012 \\
        \bottomrule
    \end{tabular}%
    }
    \label{tab:ave_std}
\end{table}

\begin{figure}
    \centering
    \includegraphics[width=0.6\textwidth]{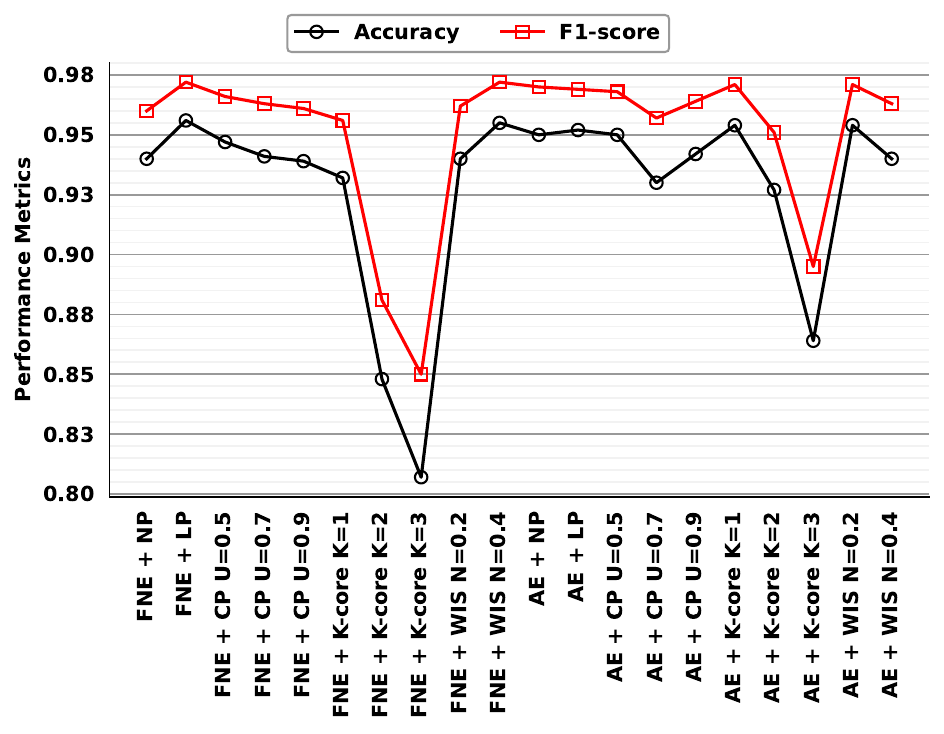}
    \caption{Performance metrics obtained through different combinations of the node embeddings and reduction techniques.}
    \label{fig:combined}
\end{figure}

In Figure~\ref{fig:combined}, each line shows the accuracy, and F1-score, for each combination of node embedding and graph reduction techniques. The attained results show that FNE combined with leaf prune and WIS with $N=0.4$ have the best performance with higher accuracy and F1-score compared to other embeddings and reduction combinations. While the cost of pruning with WIS is much more than leaf prune regarding space and time, choosing leaf prune is more rational. On the other hand for the AE, while leaf prune is still among the best, WIS with $N=0.2$ and K-core with $K=1$ have comparable performance. The first observation is that WIS is dependent on the choice of the hyper-parameter same as a K-core this makes them less stable and more time-consuming to find the best hyper-parameter value. In summary, as leaf prune is among the best for both FCG and CFG, while it can be performed in a fast and efficient manner compared to other graph reduction techniques, it should be considered as the go-to technique for graph reduction in malware detection systems.

\subsection{Explainability-based Analysis}
Explainer accuracy is used as the evaluation metric for the GNNExplainer performance. The explainer accuracy is calculated using the important subgraph determined by the explainer as the model input and measuring the model detection ability. The intuition is that the explanation subgraph should result in the same classification performance and accuracy. As explained before, GNNExplainer output is an edge mask that includes the importance weights of all the edges. To generate the subgraphs, the top $p$ percent of the edges are kept and the rest of the nodes and edges are removed.

\begin{figure}
    \captionsetup[subfigure]{justification=centering}
    \centering
    \begin{subfigure}[b]{0.25\textwidth}
        \centering
        \includegraphics[width=\textwidth]{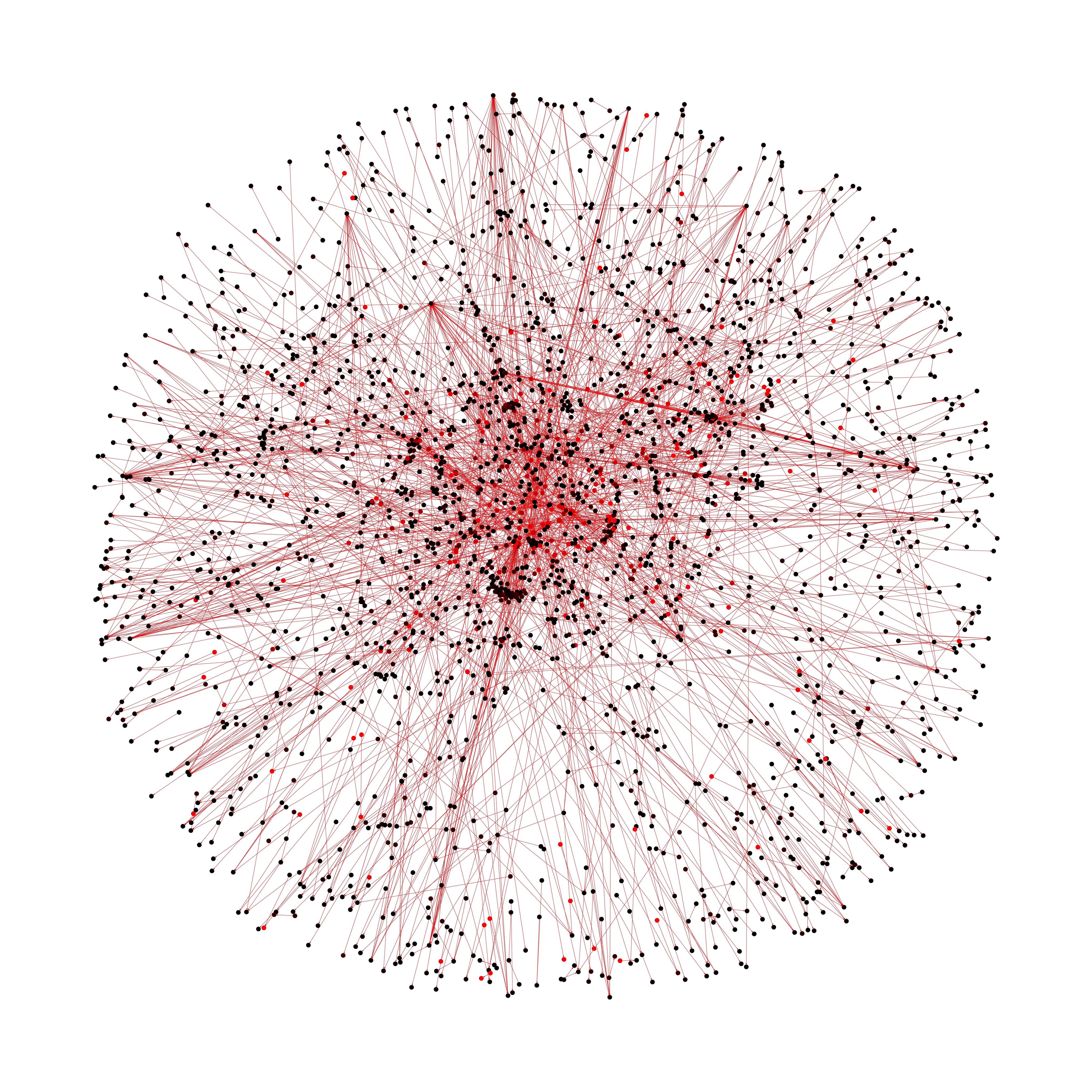}
        \caption{Important subgraph}
    \end{subfigure}
    \begin{subfigure}[b]{0.25\textwidth}
         \centering
         \includegraphics[width=\textwidth]{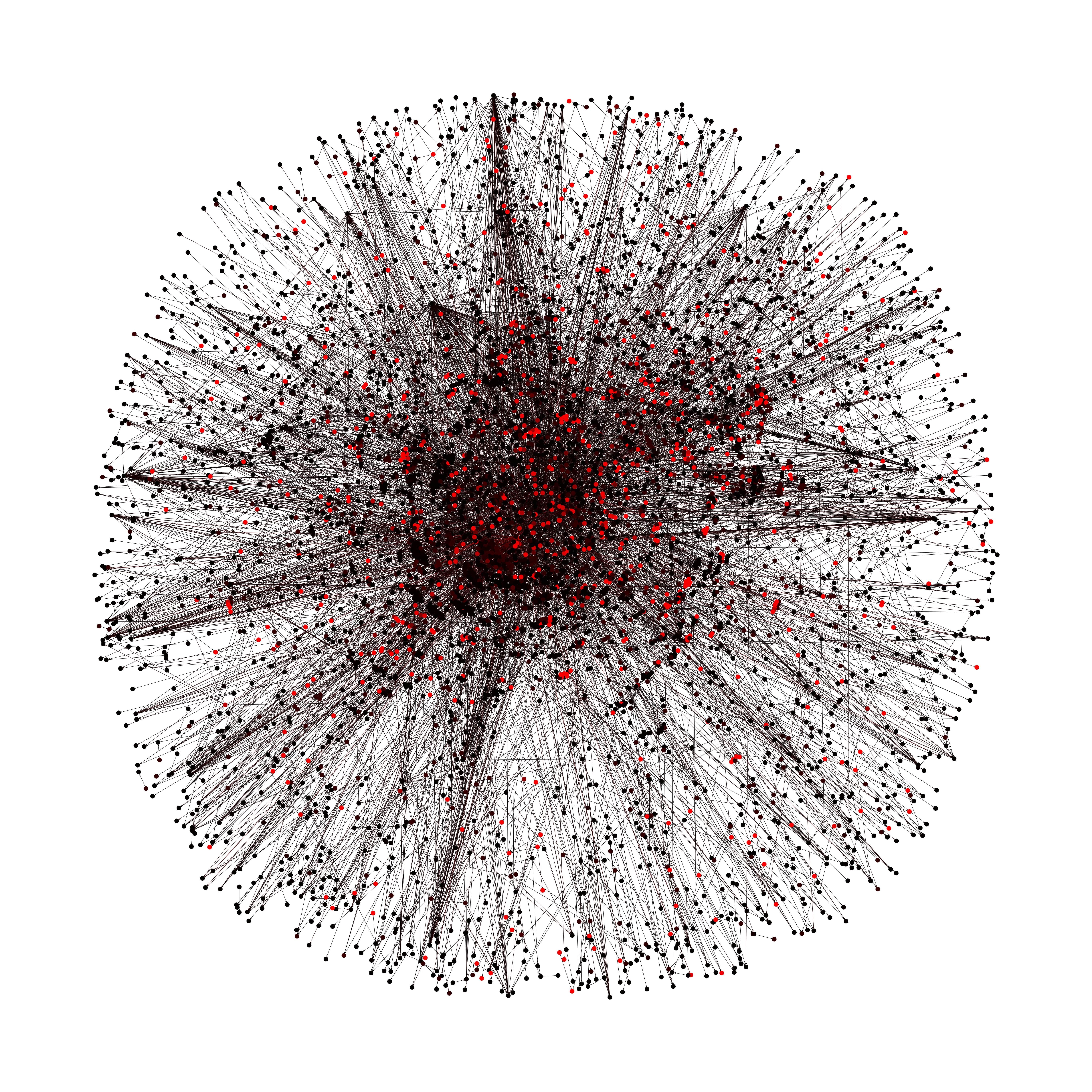}
         \caption{Unimportant subgraph}
     \end{subfigure}
     \caption{Explainer output with $p=25\%$ in panel (a)}
     \label{fig:vis_exp}
\end{figure}

\begin{figure}
    \centering
    \includegraphics[width=0.7\linewidth]{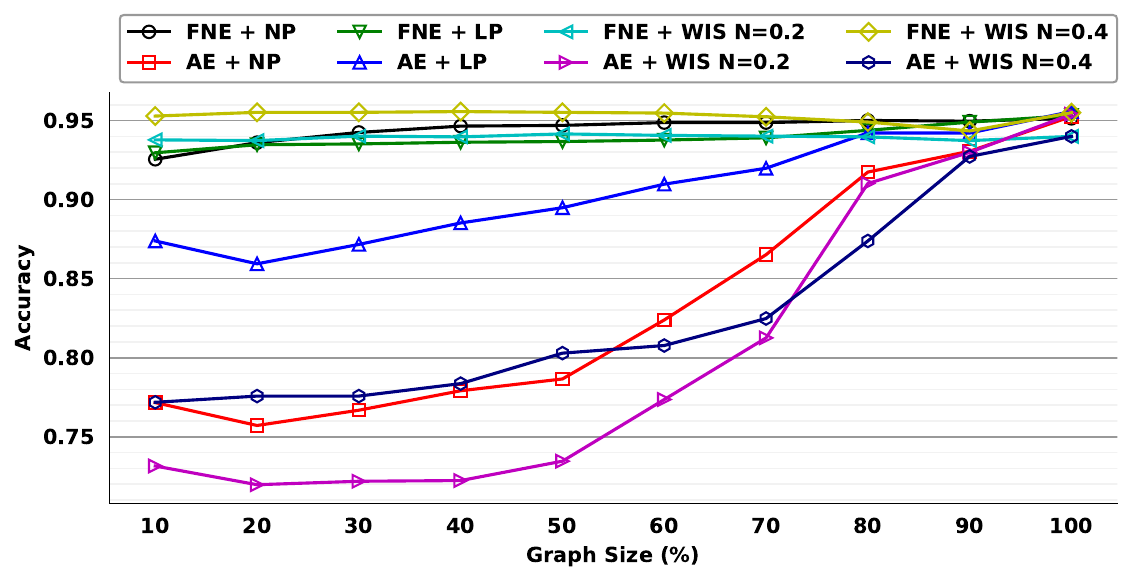}
    \caption{The classification accuracy of important subgraphs.}
    \label{fig:exp}
\end{figure}

Since leaf prune and WIS were the best reduction techniques with different node embeddings, considering both performance and efficiency, the accuracy of the explainability when we used leaf prune, WIS and no prune combined with the two node embedding techniques are presented in Figure~\ref{fig:exp}. This Figure shows the accuracy for the important subgraph with 10\% of the top edges up to 100\% with 10\% step size. Also, Figure~\ref{fig:vis_exp}, shows an example visualization of the extracted important subgraph when keep top 25\% of edges and the remaining 75\% as the unimportant subgraph.

Comparing the explainability accuracy of the FNE and AE shows that in all the cases with diffrent reduction techniques, even with small subgraphs FNE results in better explanations and achieves better accuracies starting from 92\%. On the other hand, AE results in lower accuracies, especially with WIS, which starts as low as 75\% and there is an unexpected drop from 10\% subgraphs to 20\% subgraphs, and, then, accuracy increases with bigger subgraphs as expected. One factor that may have a major effect on the explainability results is the structure of the original graph. As explained before, FNE uses FCGs, while AE uses CFGs, which are quite different and could be the source of the superiority of the GNNExplainer for FNE.

\section{Conclusion}\label{sec:conclusion}
This paper focuses on malware detection by resorting to static approaches. To do this we proposed a pipeline containing several modules including node embedding, graph reduction, graph classification, and explainability. Resorting to FCGs or CFGs for representing malware files often results in the graphs of huge sizes, which results in a very resource-intensive training and test sessions. To have a more efficient detection process, we proposed to use a structure-based graph reduction technique to make the input graph smaller and remove the nodes and edges with unimportant information. We implemented and compared several state-of-the-art graph reduction techniques to achieve this objective. Our analysis shows that among the evaluated methods, leaf prune is the most effective reduction technique, which not only reduces the size of the graphs significantly with acceptable efficiency but also improves the detection performance as well.

The other important step in the process is the technique that has been used to embed the considerable amount of information present in each node. We used two techniques called Function Name Embedding, which uses a Large Language Model to embed the function names into a continuous vector and Assembly Embedding, which embeds all the assembly instructions in each node into a 64-dimensional vector using a two-step process. The comparison between these two embedding techniques shows that in general AE outperforms FNE since it uses more actual information about the sample behaviors and instructions.

And, finally to provide more transparency about the decisions made by the model we used GNNExplainer to extract explanations, i.e., the important subgraphs with the most impact on the detection output. The obtained results show that by using a small portion (selected in a smart manner) of the graph representation of the samples, the model is still able to have an acceptable detection performance.

\bibliographystyle{elsarticle-num}
\bibliography{ref}

\begin{thebibliography}{10}
\expandafter\ifx\csname url\endcsname\relax
  \def\url#1{\texttt{#1}}\fi
\expandafter\ifx\csname urlprefix\endcsname\relax\def\urlprefix{URL }\fi
\expandafter\ifx\csname href\endcsname\relax
  \def\href#1#2{#2} \def\path#1{#1}\fi

\bibitem{CyberArk}
2024 identity security threat landscape report, \url{https://www.cyberark.com/threat-landscape/}.

\bibitem{li2019machine}
Y.~Li, K.~Xiong, T.~Chin, C.~Hu, A machine learning framework for domain generation algorithm-based malware detection, IEEE Access 7 (2019) 32765--32782.

\bibitem{pham2018static}
H.-D. Pham, T.~D. Le, T.~N. Vu, Static pe malware detection using gradient boosting decision trees algorithm, in: Future Data and Security Engineering: 5th International Conference, FDSE 2018, Ho Chi Minh City, Vietnam, November 28--30, 2018, Proceedings 5, Springer, 2018, pp. 228--236.

\bibitem{khammas2020ransomware}
B.~M. Khammas, Ransomware detection using random forest technique, ICT Express 6~(4) (2020) 325--331.

\bibitem{zhang2016malware}
J.~Zhang, Z.~Qin, H.~Yin, L.~Ou, S.~Xiao, Y.~Hu, Malware variant detection using opcode image recognition with small training sets, in: 2016 25th International Conference on Computer Communication and Networks (ICCCN), IEEE, 2016, pp. 1--9.

\bibitem{manavi2020new}
F.~Manavi, A.~Hamzeh, A new method for ransomware detection based on pe header using convolutional neural networks, in: 2020 17th international ISC conference on information security and cryptology (ISCISC), IEEE, 2020, pp. 82--87.

\bibitem{frederick2022corpus}
R.~Frederick, J.~Shapiro, R.~A. Calix, A corpus of encoded malware byte information as images for efficient classification, in: 2022 16th International Conference on Signal-Image Technology \& Internet-Based Systems (SITIS), IEEE, 2022, pp. 32--36.

\bibitem{li2022novel}
C.~Li, Q.~Lv, N.~Li, Y.~Wang, D.~Sun, Y.~Qiao, A novel deep framework for dynamic malware detection based on api sequence intrinsic features, Computers \& Security 116 (2022) 102686.

\bibitem{bensaoud2024cnn}
A.~Bensaoud, J.~Kalita, Cnn-lstm and transfer learning models for malware classification based on opcodes and api calls, Knowledge-Based Systems (2024) 111543.

\bibitem{dovom2019fuzzy}
E.~M. Dovom, A.~Azmoodeh, A.~Dehghantanha, D.~E. Newton, R.~M. Parizi, H.~Karimipour, Fuzzy pattern tree for edge malware detection and categorization in iot, Journal of Systems Architecture 97 (2019) 1--7.

\bibitem{liu2020multifamily}
X.~Liu, X.~Du, Q.~Lei, K.~Liu, Multifamily classification of android malware with a fuzzy strategy to resist polymorphic familial variants, IEEE Access 8 (2020) 156900--156914.

\bibitem{sun2021effective}
Y.~Sun, A.~K. Bashir, U.~Tariq, F.~Xiao, Effective malware detection scheme based on classified behavior graph in iiot, Ad Hoc Networks 120 (2021) 102558.

\bibitem{abusnaina2021dl}
A.~Abusnaina, M.~Abuhamad, H.~Alasmary, A.~Anwar, R.~Jang, S.~Salem, D.~Nyang, D.~Mohaisen, Dl-fhmc: Deep learning-based fine-grained hierarchical learning approach for robust malware classification, IEEE Transactions on Dependable and Secure Computing 19~(5) (2021) 3432--3447.

\bibitem{zhang2020exploring}
Y.~Zhang, X.~Chang, Y.~Lin, J.~Mi{\v{s}}i{\'c}, V.~B. Mi{\v{s}}i{\'c}, Exploring function call graph vectorization and file statistical features in malicious pe file classification, IEEE Access 8 (2020) 44652--44660.

\bibitem{cai2021learning}
M.~Cai, Y.~Jiang, C.~Gao, H.~Li, W.~Yuan, Learning features from enhanced function call graphs for android malware detection, Neurocomputing 423 (2021) 301--307.

\bibitem{wu2023iot}
C.-Y. Wu, T.~Ban, S.-M. Cheng, T.~Takahashi, D.~Inoue, Iot malware classification based on reinterpreted function-call graphs, Computers \& Security 125 (2023) 103060.

\bibitem{amer2020dynamic}
E.~Amer, I.~Zelinka, A dynamic windows malware detection and prediction method based on contextual understanding of api call sequence, Computers \& Security 92 (2020) 101760.

\bibitem{amer2021multi}
E.~Amer, I.~Zelinka, S.~El-Sappagh, A multi-perspective malware detection approach through behavioral fusion of api call sequence, Computers \& Security 110 (2021) 102449.

\bibitem{li2022dmalnet}
C.~Li, Z.~Cheng, H.~Zhu, L.~Wang, Q.~Lv, Y.~Wang, N.~Li, D.~Sun, Dmalnet: Dynamic malware analysis based on api feature engineering and graph learning, Computers \& Security 122 (2022) 102872.

\bibitem{gao2022malware}
Y.~Gao, H.~Hasegawa, Y.~Yamaguchi, H.~Shimada, Malware detection by control-flow graph level representation learning with graph isomorphism network, IEEE Access 10 (2022) 111830--111841.

\bibitem{nguyen2018auto}
M.~H. Nguyen, D.~Le~Nguyen, X.~M. Nguyen, T.~T. Quan, Auto-detection of sophisticated malware using lazy-binding control flow graph and deep learning, Computers \& Security 76 (2018) 128--155.

\bibitem{herath2022cfgexplainer}
J.~D. Herath, P.~P. Wakodikar, P.~Yang, G.~Yan, Cfgexplainer: Explaining graph neural network-based malware classification from control flow graphs, in: 2022 52nd Annual IEEE/IFIP International Conference on Dependable Systems and Networks (DSN), IEEE, 2022, pp. 172--184.

\bibitem{gao2021gdroid}
H.~Gao, S.~Cheng, W.~Zhang, Gdroid: Android malware detection and classification with graph convolutional network, Computers \& Security 106 (2021) 102264.

\bibitem{bruna2013spectral}
J.~Bruna, W.~Zaremba, A.~Szlam, Y.~LeCun, Spectral networks and locally connected networks on graphs, arXiv preprint arXiv:1312.6203 (2013).

\bibitem{kipf2016semi}
T.~N. Kipf, M.~Welling, Semi-supervised classification with graph convolutional networks, arXiv preprint arXiv:1609.02907 (2016).

\bibitem{hashemi2024comprehensive}
M.~Hashemi, S.~Gong, J.~Ni, W.~Fan, B.~A. Prakash, W.~Jin, A comprehensive survey on graph reduction: Sparsification, coarsening, and condensation, arXiv preprint arXiv:2402.03358v4 (2024).

\bibitem{hu2013survey}
P.~Hu, W.~C. Lau, A survey and taxonomy of graph sampling, arXiv preprint arXiv:1308.5865 (2013).

\bibitem{li2023attend}
X.~Li, K.~Wang, H.~Deng, Y.~Liang, D.~Wu, Attend who is weak: Enhancing graph condensation via cross-free adversarial training, arXiv preprint arXiv:2311.15772 (2023).

\bibitem{razin2023abilitygraphneuralnetworks}
N.~Razin, T.~Verbin, N.~Cohen, \href{https://arxiv.org/abs/2211.16494}{On the ability of graph neural networks to model interactions between vertices} (2023).
\newblock \href {http://arxiv.org/abs/2211.16494} {\path{arXiv:2211.16494}}.
\newline\urlprefix\url{https://arxiv.org/abs/2211.16494}

\bibitem{warmsley2022survey}
D.~Warmsley, A.~Waagen, J.~Xu, Z.~Liu, H.~Tong, A survey of explainable graph neural networks for cyber malware analysis, in: 2022 IEEE International Conference on Big Data (Big Data), IEEE, 2022, pp. 2932--2939.

\bibitem{arp2014drebin}
D.~Arp, M.~Spreitzenbarth, M.~Hubner, H.~Gascon, K.~Rieck, C.~Siemens, Drebin: Effective and explainable detection of android malware in your pocket., in: Ndss, Vol.~14, 2014, pp. 23--26.

\bibitem{kinkead2021towards}
M.~Kinkead, S.~Millar, N.~McLaughlin, P.~O’Kane, Towards explainable cnns for android malware detection, Procedia Computer Science 184 (2021) 959--965.

\bibitem{ullah2022explainable}
F.~Ullah, A.~Alsirhani, M.~M. Alshahrani, A.~Alomari, H.~Naeem, S.~A. Shah, Explainable malware detection system using transformers-based transfer learning and multi-model visual representation, Sensors 22~(18) (2022) 6766.

\bibitem{alani2023xmal}
M.~M. Alani, A.~Mashatan, A.~Miri, Xmal: A lightweight memory-based explainable obfuscated-malware detector, Computers \& Security 133 (2023) 103409.

\bibitem{ying2019gnnexplainer}
Z.~Ying, D.~Bourgeois, J.~You, M.~Zitnik, J.~Leskovec, Gnnexplainer: Generating explanations for graph neural networks, Advances in neural information processing systems 32 (2019).

\bibitem{luo2020parameterized}
D.~Luo, W.~Cheng, D.~Xu, W.~Yu, B.~Zong, H.~Chen, X.~Zhang, Parameterized explainer for graph neural network, Advances in neural information processing systems 33 (2020) 19620--19631.

\bibitem{yuan2021explainability}
H.~Yuan, H.~Yu, J.~Wang, K.~Li, S.~Ji, On explainability of graph neural networks via subgraph explorations, in: International conference on machine learning, PMLR, 2021, pp. 12241--12252.

\bibitem{domingos2012few}
P.~Domingos, A few useful things to know about machine learning, Communications of the ACM 55~(10) (2012) 78--87.

\bibitem{halevy2009unreasonable}
A.~Halevy, P.~Norvig, F.~Pereira, The unreasonable effectiveness of data, IEEE intelligent systems 24~(2) (2009) 8--12.

\bibitem{geiger2020garbage}
R.~S. Geiger, K.~Yu, Y.~Yang, M.~Dai, J.~Qiu, R.~Tang, J.~Huang, Garbage in, garbage out? do machine learning application papers in social computing report where human-labeled training data comes from?, in: Proceedings of the 2020 conference on fairness, accountability, and transparency, 2020, pp. 325--336.

\bibitem{Roberts2011}
J.-M. Roberts, Virus share (2011).

\bibitem{Quist2009}
D.~Quist, Open malware (2009).

\bibitem{baecher2006nepenthes}
P.~Baecher, M.~Koetter, T.~Holz, M.~Dornseif, F.~Freiling, The nepenthes platform: An efficient approach to collect malware, in: Recent Advances in Intrusion Detection: 9th International Symposium, RAID 2006 Hamburg, Germany, September 20-22, 2006 Proceedings 9, Springer, 2006, pp. 165--184.

\bibitem{shoshitaishvili2016state}
Y.~Shoshitaishvili, R.~Wang, C.~Salls, N.~Stephens, M.~Polino, A.~Dutcher, J.~Grosen, S.~Feng, C.~Hauser, C.~Kruegel, G.~Vigna, Sok: (state of) the art of war: Offensive techniques in binary analysis (2016).

\bibitem{stephens2016driller}
N.~Stephens, J.~Grosen, C.~Salls, A.~Dutcher, R.~Wang, J.~Corbetta, Y.~Shoshitaishvili, C.~Kruegel, G.~Vigna, Driller: Augmenting fuzzing through selective symbolic execution (2016).

\bibitem{shoshitaishvili2015firmalice}
Y.~Shoshitaishvili, R.~Wang, C.~Hauser, C.~Kruegel, G.~Vigna, Firmalice - automatic detection of authentication bypass vulnerabilities in binary firmware (2015).

\bibitem{wang2020minilm}
W.~Wang, F.~Wei, L.~Dong, H.~Bao, N.~Yang, M.~Zhou, Minilm: Deep self-attention distillation for task-agnostic compression of pre-trained transformers, Advances in Neural Information Processing Systems 33 (2020) 5776--5788.

\bibitem{peng2024malgne}
H.~Peng, J.~Yang, D.~Zhao, X.~Xu, Y.~Pu, J.~Han, X.~Yang, M.~Zhong, S.~Ji, Malgne: Enhancing the performance and efficiency of cfg-based malware detector by graph node embedding in low dimension space, IEEE Transactions on Information Forensics and Security (2024).

\bibitem{razin2023ability}
N.~Razin, T.~Verbin, N.~Cohen, On the ability of graph neural networks to model interactions between vertices, Advances in Neural Information Processing Systems 36 (2023) 26501--26545.

\bibitem{paszke2017automatic}
A.~Paszke, S.~Gross, S.~Chintala, G.~Chanan, E.~Yang, Z.~DeVito, Z.~Lin, A.~Desmaison, L.~Antiga, A.~Lerer, Automatic differentiation in pytorch (2017).

\bibitem{DBLP:journals/corr/abs-1903-02428}
M.~Fey, J.~E. Lenssen, \href{http://arxiv.org/abs/1903.02428}{Fast graph representation learning with pytorch geometric}, CoRR abs/1903.02428 (2019).
\newblock \href {http://arxiv.org/abs/1903.02428} {\path{arXiv:1903.02428}}.
\newline\urlprefix\url{http://arxiv.org/abs/1903.02428}

\bibitem{paper:hagberg:2008}
A.~A. Hagberg, D.~A. Schult, P.~J. Swart, \href{http://conference.scipy.org/proceedings/SciPy2008/paper_2/}{Exploring network structure, dynamics, and function using networkx}, in: G.~Varoquaux, T.~Vaught, J.~Millman (Eds.), Proceedings of the 7th Python in Science Conference, Pasadena, CA USA, 2008, pp. 11 -- 15.
\newline\urlprefix\url{http://conference.scipy.org/proceedings/SciPy2008/paper_2/}

\bibitem{yang2021bodmas}
L.~Yang, A.~Ciptadi, I.~Laziuk, A.~Ahmadzadeh, G.~Wang, Bodmas: An open dataset for learning based temporal analysis of pe malware, in: 2021 IEEE Security and Privacy Workshops (SPW), IEEE, 2021, pp. 78--84.

\bibitem{dikedataset}
G.-A. Iosif, Dikedataset, \url{https://github.com/iosifache/DikeDataset}, accessed on February 27, 2024 (2021).

\bibitem{practicalsecurity2024pe}
{Practical Security Analytics LLC}, Pe malware machine learning dataset, \url{https://practicalsecurityanalytics.com/pe-malware-machine-learning-dataset/}, accessed: 2024-08-06 (2024).

\end{thebibliography}

\end{document}